\documentclass[useAMS,usenatbib]{mn2e}
 \usepackage{graphicx}
 \usepackage{amsmath}
\usepackage{txfonts}
\usepackage[T1]{fontenc}
\usepackage{color}

 \title[Variability of wind absorption lines in LMXBs]
{Variable Doppler shifts of the thermal wind absorption lines in low-mass X-ray binaries}

\author [Madej et al.]
{O.K. Madej$^{1,2}$\thanks{E-mail: O.Madej@sron.nl}, P.G. Jonker$^{2,1,3}$, M. D\'{i}az Trigo$^{4}$, I. Mi\v{s}kovi\v{c}ov\'{a}$^{5,6}$\\
\\
\normalsize{$^{1}$Department of Astrophysics/IMAPP, Radboud University Nijmegen, P.O. Box 9010, 6500 GL Nijmegen, The Netherlands}\\
\normalsize{$^{2}$SRON Netherlands Institute for Space Research, Sorbonnelaan 2, 3584 CA Utrecht, The Netherlands}\\
\normalsize{$^{3}$Harvard-Smithsonian Center for Astrophysics, 60 Garden Street, Cambridge, MA 02138, USA}\\
\normalsize{$^{4}$ESO, Karl-Schwarzschild-Strasse 2, 85748 Garching bei M\"{u}nchen, Germany }\\
\normalsize{$^{5}$Dr. Karl Remeis-Sternwarte and Erlangen Center for Astroparticle Physics, Universit\"{a}t Erlangen-N\"{u}rnberg,} \\
\normalsize{Sternwartstr. 7, 96049 Bamberg, Germany }\\
\normalsize{$^{6}$Czech Technical University, Faculty of Electrical Engineering, Technick\'a 2, 160 00 Praha 6, Czech Republic }}

\begin{document}



\maketitle
\label{firstpage}
\def\apjl{ApJ}
\def\aj{AJ}
\def\apj{ApJ}
\def\pasp{PASP}
\def\spie{SPIE}
\def\apjs{ApJS}
\def\araa{ARAA}
\def\aap{A\&A}
\def\nat{Nature}
\def\mnras{MNRAS}
\def\prd{Phys.Rev.D}

\begin{abstract}
In this paper we address the general applicability of the method pioneered by \citet{Zhang2012} in which the motion of the compact object can be tracked using wind X-ray absorption lines. We present the velocity measurements of the thermal wind lines observed in the X-ray spectrum of a few low-mass X-ray binaries: GX 13+1, H 1743$-$322, GRO J1655$-$40 and GRS 1915+105. We find that the variability in the velocity of the wind lines in about all of the sources is larger than conceivable radial velocity variations of the compact object. GX 13+1 provides a potential exception, although it would require the red giant star to be massive with a mass of $\approx 5-6\ M_{\odot}$. We conclude that the variability of the source luminosity occurring on a time scale of days/months can affect the outflow properties making it difficult to track the orbital motion of the compact object using current observations. Given the intrinsic variability of the outflows we suggest that low-mass X-ray binaries showing stable coronae instead of an outflow (e.g. 4U 1254$-$69, MXB 1659$-$29, 4U 1624$-$49) could be more suitable targets for tracking the orbital motion of the compact object.

\end{abstract}


\section{Introduction}

It has been shown by \citet{Zhang2012} that the disc outflow lines in GRO J1655$-$40 and LMC X-3 track the motion of the compact object around the binary center of mass. The authors  point out that since the disc outflow moves together with the accretion disc the Doppler motion of the wind absorption lines can be considered as that of the compact object. If this method of tracking the orbital motion of the disc and compact object is universally applicable then it could prove useful in measuring the masses of the black holes/neutron stars in X-ray binaries.\\
So far, masses of the compact objects in X-ray binaries have been measured using the Kepler's third law expressed in the form of the mass function. The mass function can be solved for the mass of the compact object using phase-resolved optical spectroscopy showing e.g. Doppler motion of the absorption lines originating at the surface of the donor and measuring the Doppler broadening of the absorption lines providing the mass ratio, as well as the binary inclination \citep[for more details see e.g.][]{Ratti2012}. However, since only the motion of the donor star is detected directly in these systems, the resulting mass measurement of the compact object can be model dependent. For instance corotation of the mass donor rotation around its axis and the orbital motion are assumed. Similarly, the ellipsoidal modulations used to obtain the inclination often depend on modeling the residual accretion disc contribution to the optical and near-infrared light. In order to break this dependence, there have been attempts to measure the motion of the compact object itself using the optical emission line features originating in the disc \citep{Soria1998}. In this way a single-lined spectroscopic binary would become a `double'-lined spectroscopic binary in which case the mass ratio can be measured directly. These lines are, however, usually asymmetric and variable which makes it difficult to model them correctly and determine the compact object motion reliably. As shown by \citet{Zhang2012} detecting narrow absorption lines in X-rays originating in the disc outflow instead provides a much better opportunity to track the motion of the compact object. \\
\begin{table}
\begin{center}
\caption{Log of the {\it Chandra} HETGS observations. }
\begin{tabular}{lcccc}
\hline
\hline
 \#&Obs. ID  & Mode$^{*}$ &  Date & $T_{\rm exp}$  [ks]\\
\hline 
GX 13+1 & & & & \\
1& 2708 & TE-G & 08/10/2002 & 29 \\
 2& 11815 & TE-F & 24/07/2010 & 28 \\
 3&11816 & TE-F & 30/07/2010 & 28 \\
 4&11814 & TE-F & 01/08/2010 & 28 \\
 5&11817 & TE-F & 03/08/2010 & 28 \\
 6&11818 & CC-F & 05/08/2010 & 23 \\
 7&13197 & CC-G & 17/02/2011 & 10 \\
\hline
H 1743-322  & & & & \\
&  3803 & CC-G &  01/05/2003 & 48 \\
& 3805 & CC-G & 23/06/2003 & 50 \\
& 3806 & CC-G & 30/07/2003 & 50 \\
\hline
GRO J1655$-$40  & & & & \\
& 5460 & CC-G & 12/03/2005 & 24\\
& 5461$^{**}$ & CC-G & 01/04/2005 & 44 \\
\hline
GRS 1915+105  & & & & \\
& 6579 & CC-F & 01/12/2005 & 12 \\
& 6580 & CC-F & 01/12/2005 & 12 \\
& 6581 & CC-F & 03/12/2005 & 10 \\
& 7485 & TE-G & 14/08/2007 & 47 \\
\hline
\end{tabular} 
\end{center}
{\footnotesize $^{*}$ Modes: Time Graded (TE-G), Time Faint (TE-F), Continuous Clocking Faint (CC-F), Continuous Clocking Graded (CC-G)\\
$^{**}$ this observation shows a wind with many absorption lines and the column density which is higher than typically observed in LMXBs}
\vspace{-3mm}
\end{table}
Absorption lines originating from highly ionized disc outflows have been frequently found in the X-ray spectra of black hole X-ray binaries (XRBs). One of the clear properties of these outflows is that they are observed only in high-inclination (> 60$^{\circ}$) black hole XRBs and most of the time when the sources are in the soft (spectrum) state with a high X-ray flux \citep{Ponti2012}. It is therefore likely that these winds have an equatorial geometry and opening angles of few tens of degrees. \\
In this study we consider a few neutron star/black hole low-mass X-ray binaries (LMXBs) showing disc outflows: GX 13+1, H 1743$-$322, GRO J1655$-$40 and GRS 1915+105. We note that none of these sources show a strong stellar wind which could complicate the study of the disc outflow. The selected LMXBs were observed at least two times with {\it Chanda's} High Energy Transmission Grating (HETGS) and all the spectra considered in the analysis contain strong outflow absorption lines.\\ 
GX 13+1 consists of a neutron star and an evolved late-type K5 III star \citep{Fleischman1985,Bandyopadhyay1999}. The X-ray spectrum of the source is characterized by the presence of a soft thermal component and high flux. GX 13+1 is likely a high-inclination source \citep[60$^{\circ}$<i<80$^{\circ}$, ][]{Diaz2012}. \citet{Ueda2004} detected outflow lines e.g. Fe XXVI, Fe XXV, S XVI, Si XIV with a velocity of $-400$ ${\rm km\ s}^{-1}$ in {\it Chandra} HETGS observation of this source. The properties of the detected lines are consistent with a wind driven by thermal/radiative pressure. We note that this source is one of the very few neutron star LMXBs showing a thermal outflow. The absorbers found in most of the dipping neutron star LMXBs appear to be in the form of a steady corona: they are not showing signs of an outflow \citep{DiazBoirin2012}. \\
H 1743$-$322, GRO J1655$-$40 and GRS 1915+105 likely contain a black hole. These sources have a high inclination (i>60$^{\circ}$) and show prominent disc outflows in the soft states \citep{Ponti2012}. Most of the {\it Chandra} outflow observations show narrow Fe XXVI and Fe XXV absorption lines and the properties of these outflows inferred from the observations are consistent with them being driven by a thermal/radiative mechanism \citep{Miller2006a,Neilsen2009,Ueda2009}. On one occasion, however, GRO J1655$-$40 showed an outflow with a high column density and a spectrum that contains many absorption lines. It has been suggested by \citet{Miller2006b} that the detected wind must be driven by magnetic rather than thermal pressure. This observation of GRO J1655$-$40 was used by \citet{Zhang2012} in order to measure the motion of the black hole around the binary center of mass.\\
In this paper we address the applicability of the method of tracking the motion of the compact object using outflow absorption lines. We analyze the {\it Chandra} observations of a sample of LMXBs focusing on the variability in the velocity of the outflow lines. Finally, we discuss the possible cause of this variability and conclude that the detected variability is unlikely to be caused by the motion of the compact object around the center of mass of the binary in all sources. Our findings suggest that only symmetric and stable winds can be used to track the motion of the compact object. These characteristic of the wind, however, do not appear to be common in the outflows observed in LMXB.
\begin{figure*} 
\vspace{-3mm}
\begin{tabular}{lll}
\includegraphics[height=0.2\textwidth]{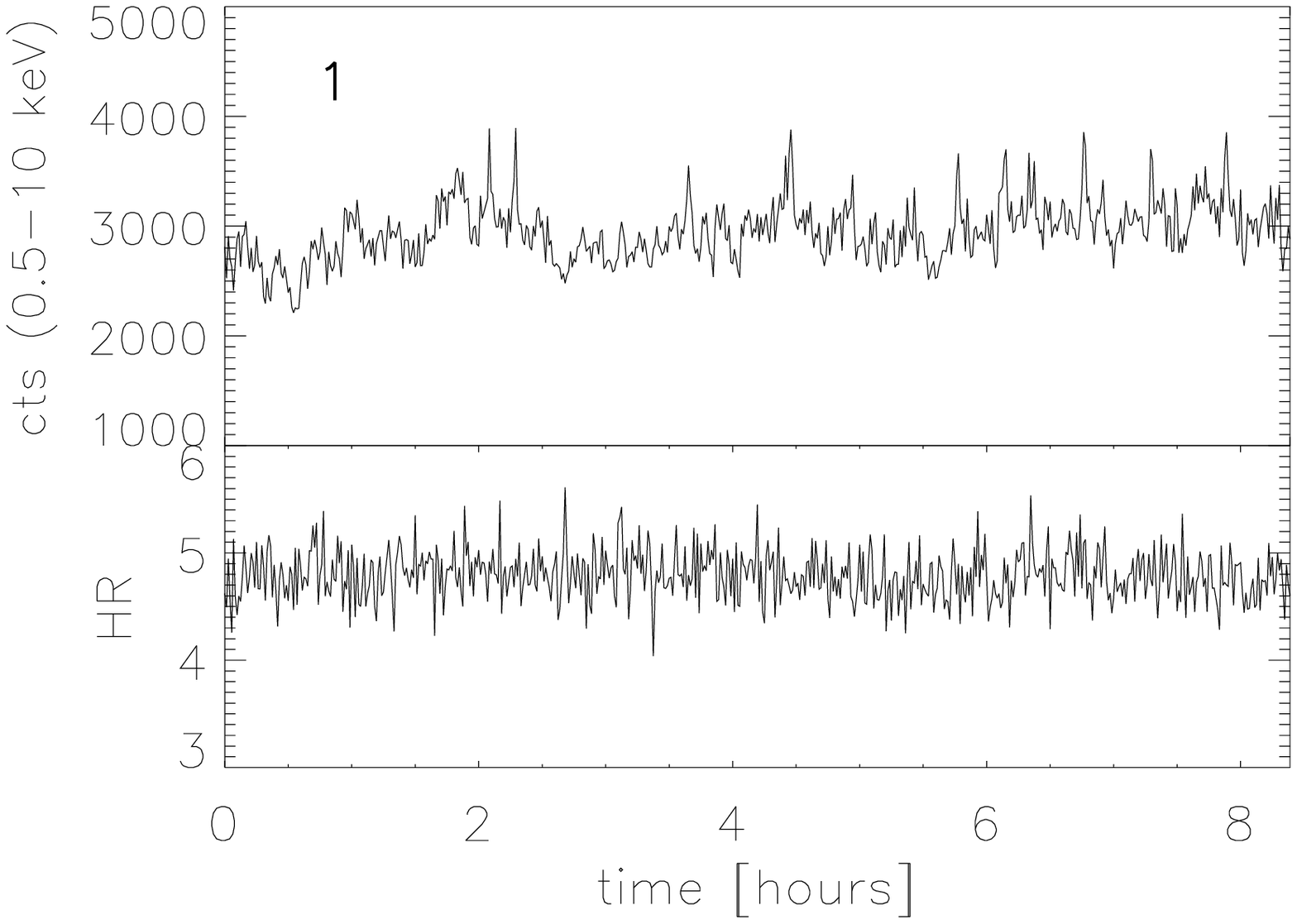}\includegraphics[height=0.2\textwidth]{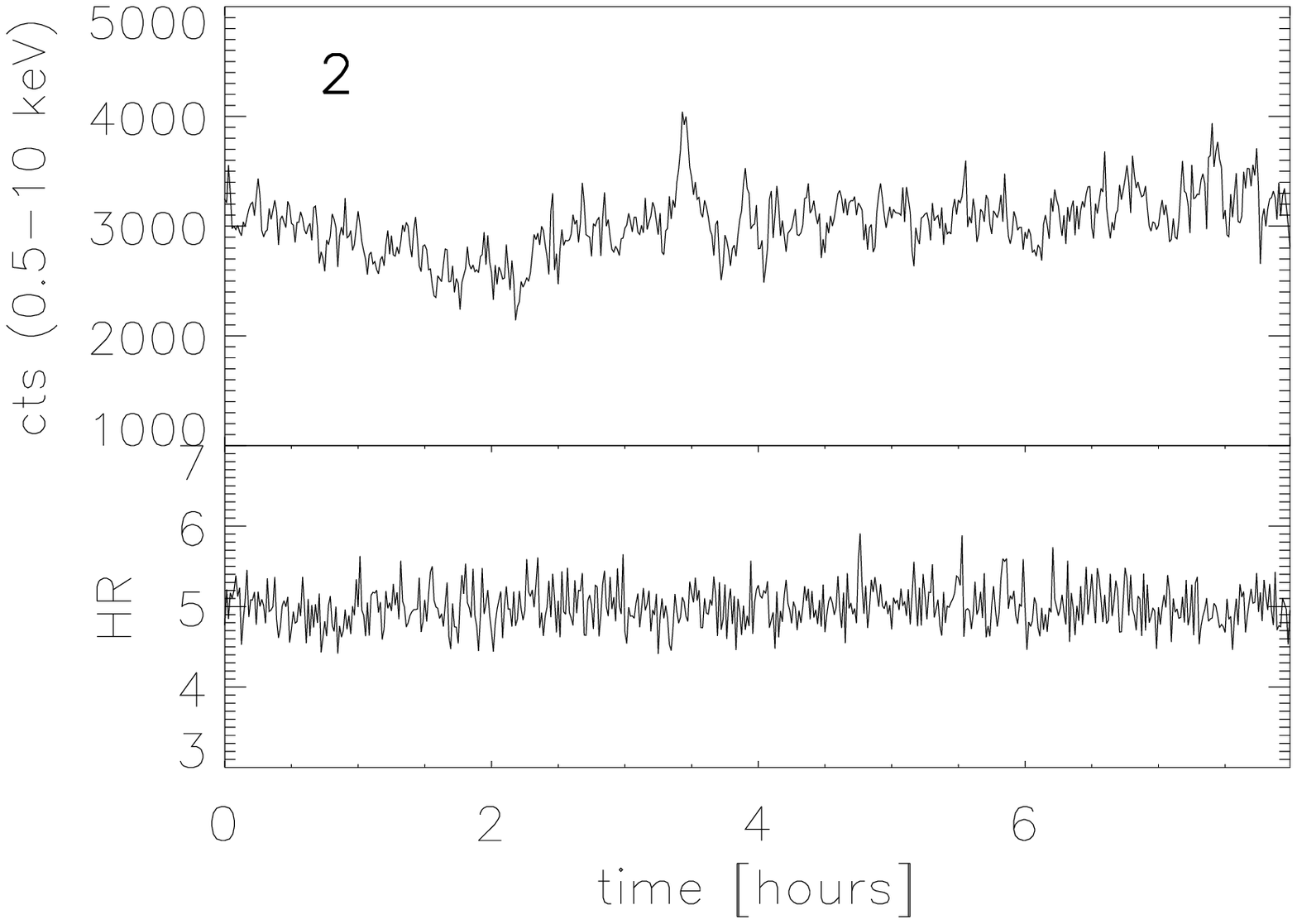}\includegraphics[height=0.2\textwidth]{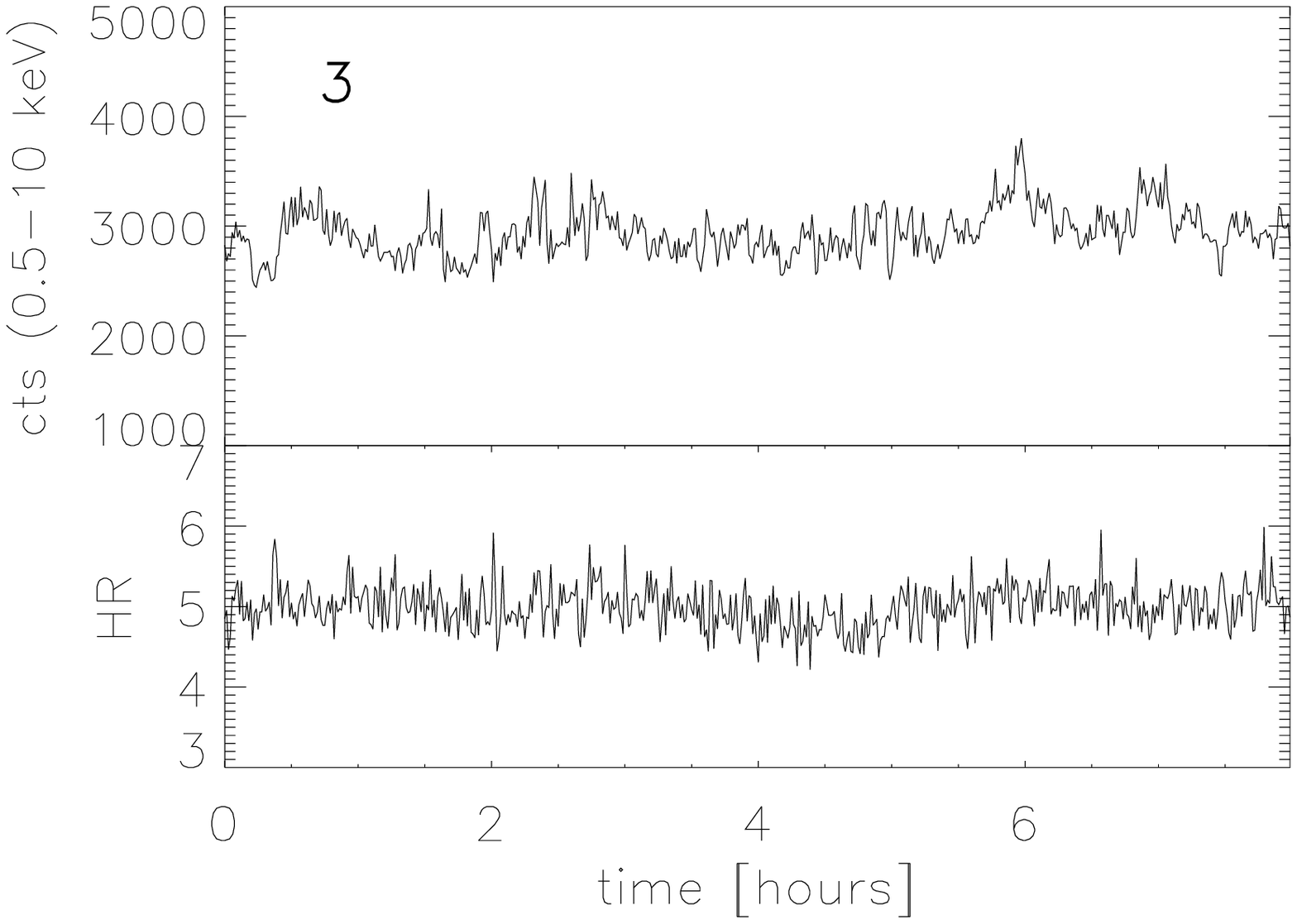}\\
\includegraphics[height=0.2\textwidth]{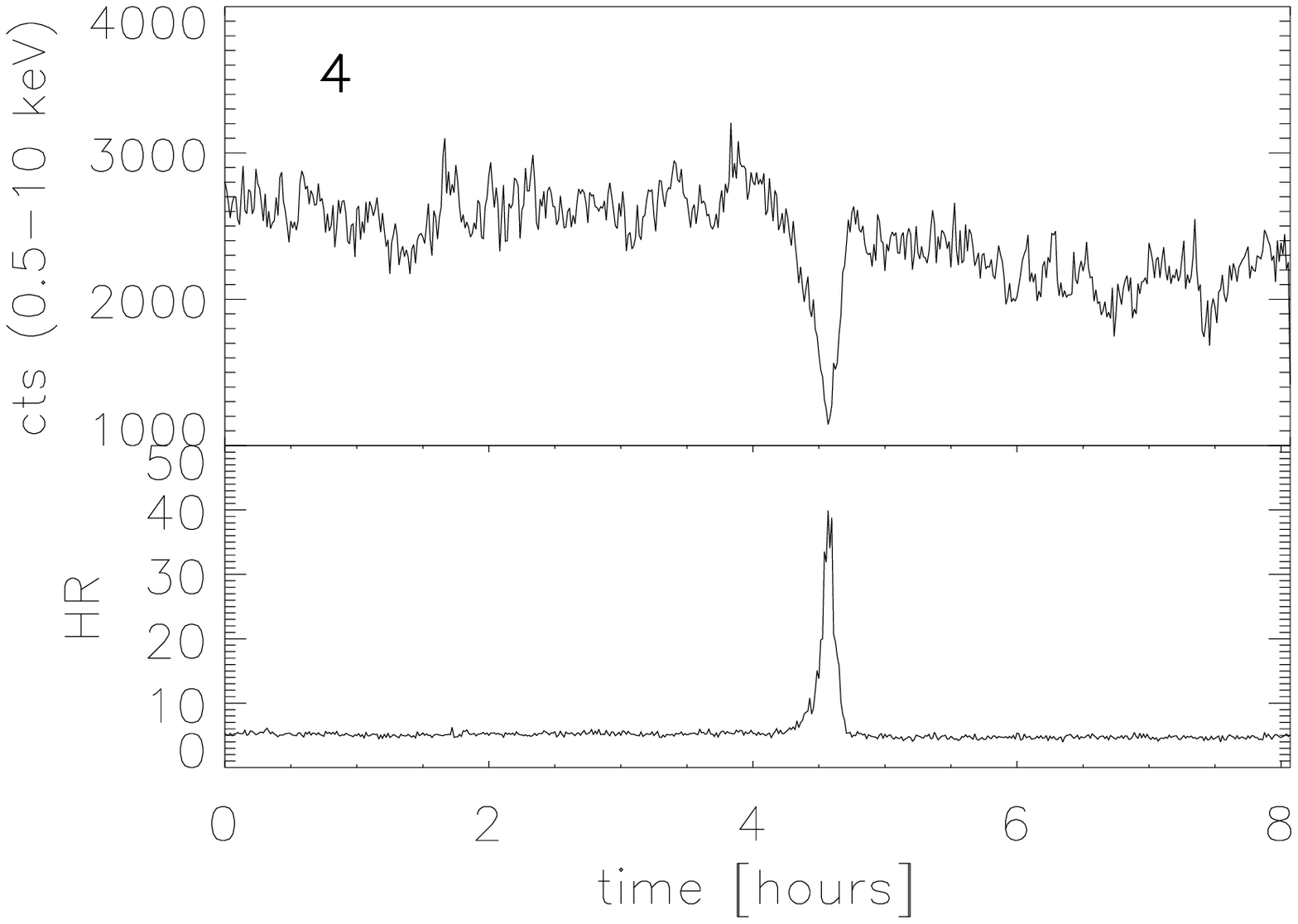}\includegraphics[height=0.2\textwidth]{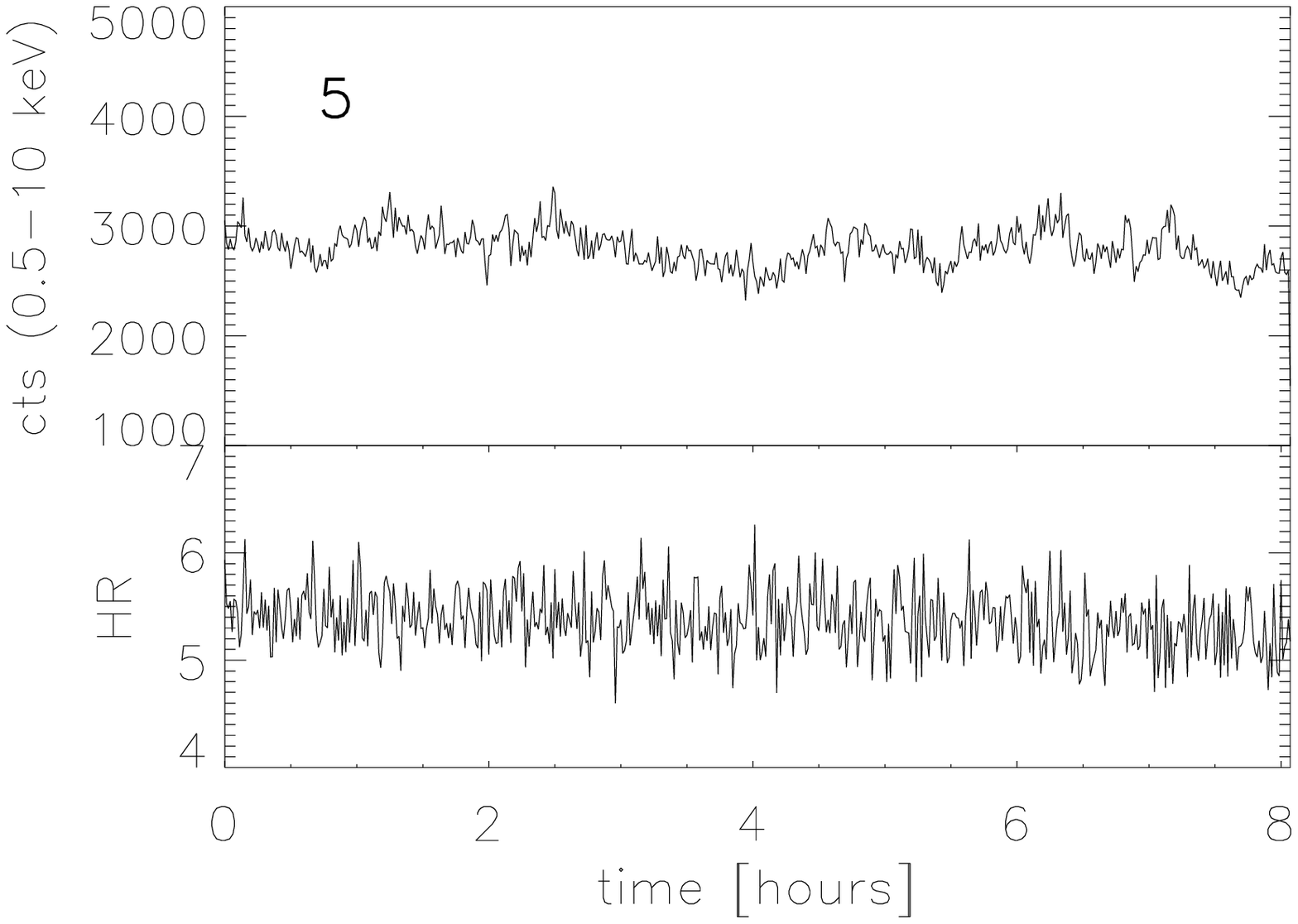}\includegraphics[height=0.2\textwidth]{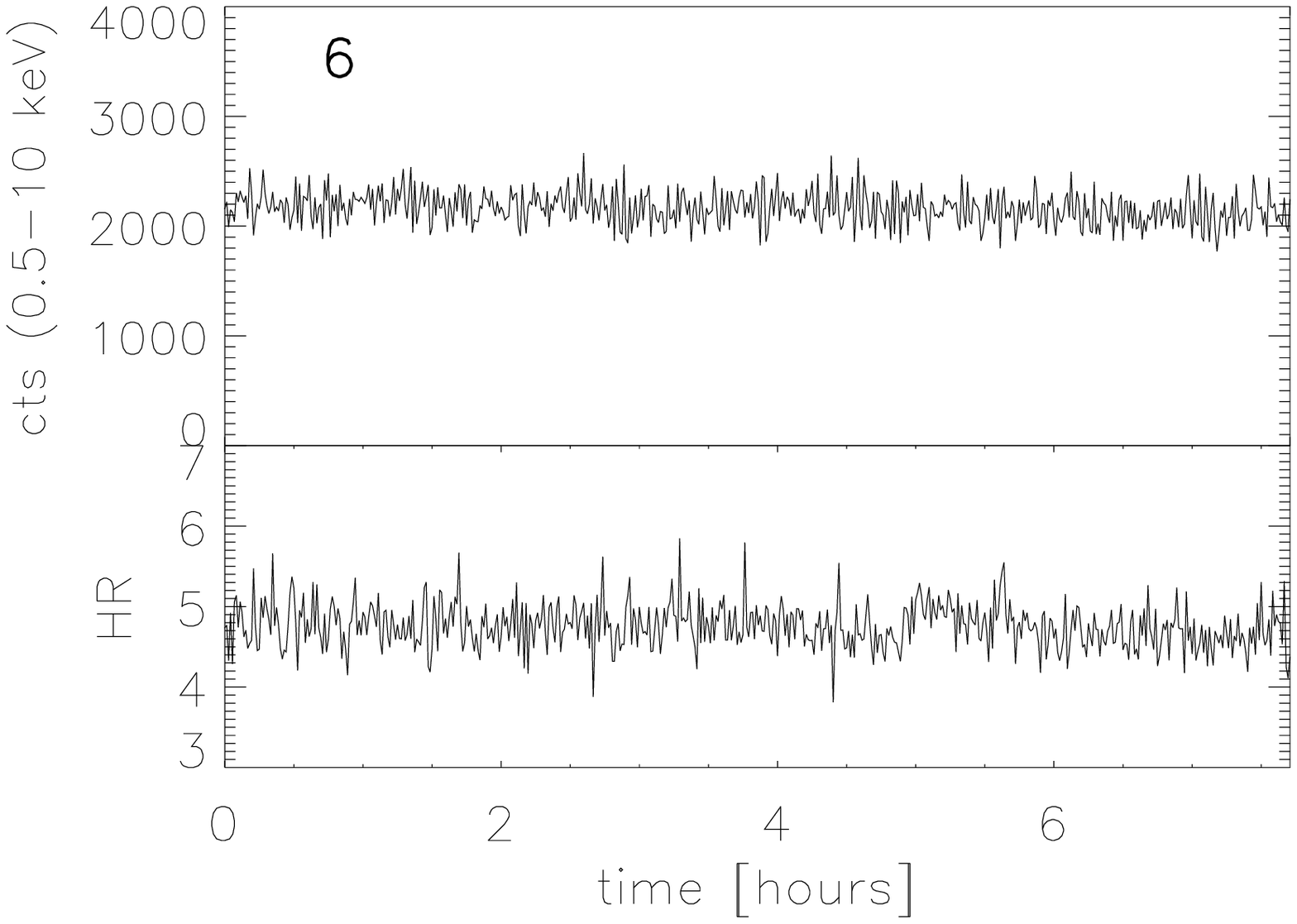}\\
\includegraphics[height=0.2\textwidth]{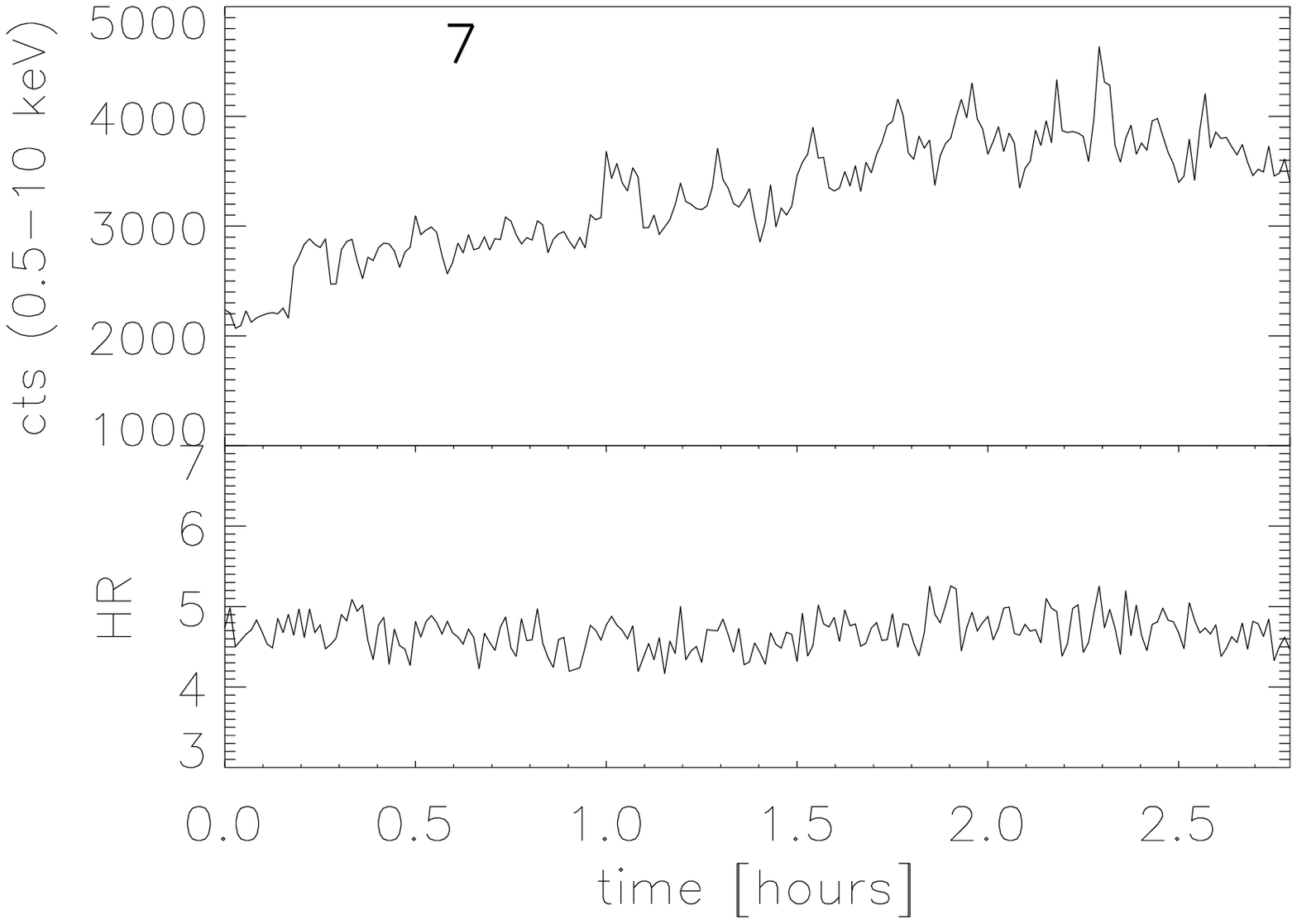}\\
\vspace{-3mm}
\end{tabular}
 \caption{MEG and HEG light curves of the LMXB GX 13+1 extracted using $+1, -1$ order MEG and HEG events. The hardness ratio is calculated using the counts in the energy range $2-10$ keV divided by the counts in the energy range $0.5-2$ keV. Note a dip in the light curve of obs. 4, which corresponds to significant spectral hardening. } 
\end{figure*}
\section{Observations and data reduction}
GX 13+1, GRS 1915+105, H 1743$-$322 and GRO J1655$-$40 were observed by {\it Chandra} HETGS between 2002 and 2011 (see Table 1). We selected only the observations showing prominent outflow lines. The observations were performed either in the Time Exposure (TE) with the read out mode set to Faint or Graded or in the Continuous Clocking (CC) with the read out mode Faint or Graded. \\
We reduce the data using standard tools provided by the {\sc CIAO} software. Since the zeroth order in the TE observations is significantly piled-up we use the {\sc tg$\_$findzo} tool in order to determine the correct position of the source. The center of the source is determined by computing the intersection of the data dispersed along one of the grating arms and the readout streak that is associated with the zeroth order image of the source. The center can be determined with an error of less than 0.1 pixel which corresponds to 0.000556 \AA \footnotemark[1] \footnotetext[1]{http://cxc.cfa.harvard.edu/cdo/about\_chandra} ($\approx 80\ {\rm km\ s^{-1}}$ at 2\ \AA). In the case of the observations taken in the CC mode we check that the position of the spectral lines in the positive and negative order are consistent within the 3 $\sigma$ errorbar. In the case that the position of the spectral lines in the positive and negative order do not match we correct the position of the zeroth order during the extraction of the spectra. Additionally, in the CC mode we apply the {\sc destreak} function in order to remove the read-out streak. The default High Energy Grating (HEG) mask used for the spectral extraction has a width of 73 pixels in the cross-dispersion direction. However, in order to improve the HEG flux determination at short wavelength we use a narrower mask of 26 pixels. The source is highly absorbed by the interstellar medium, therefore we extract only the first order HEG spectra. The HEG spectrum has also a resolution twice as good as a Medium Energy Grating (MEG) spectrum and as a result it is a factor of $\approx2$ less affected by pile-up than a MEG spectrum. During the analysis the positive and negative order spectra are fitted simultaneously. Additionally the barycentric correction is applied to all the event files using the {\sc axbary} function available in {\sc CIAO}. The duration of the analyzed observations as a fraction of the {\it Chandra} orbital period varies between 0.04 and 0.2. As a result the error on the position of the absorption line in the analyzed spectra introduced by applying the barycentric correction to the average spectrum is small compared to the statistical error.\\ 
The response and effective area files are created using the standard {\sc CIAO} tools. All HEG spectra are binned by a factor of 2 ($\approx$1/3 of the first order FWHM$_{\rm HEG}$=0.012\ \AA). The number of counts per bin exceeds $\approx15-20$ counts, hence we use $\chi^2$ statistics to fit the data. The spectra are analyzed using the  {\sc isis}\footnotemark[2] \footnotetext[2]{http://space.mit.edu/cxc/isis} and {\sc spex}\footnotemark[3] \footnotetext[3]{http://www.sron.nl/spex}  software packages. The errors on the fit parameters reported throughout this paper are calculated for $\Delta\chi^2 = 1.0$ ($1\sigma$ for a single parameter).\\
We extract the MEG and HEG light curves in the energy range 0.5-10 keV using $+1, -1$ order MEG and HEG events. We use time bins of 50 s long for each light curve. Additionally the hardness ratio $HR={\rm Counts}(2-10\ {\rm keV})/{\rm Counts}(0.5-2\ {\rm keV})$ for all observations is calculated (see Fig. 1). The light curves are not corrected for pile-up. We notice a dip in the light curves of observation 4. Hence, during the spectral extraction we exclude the part of the observation where the dip occurs. 
\begin{figure*}  
\vspace{-3mm}
\includegraphics[width=0.85\textwidth]{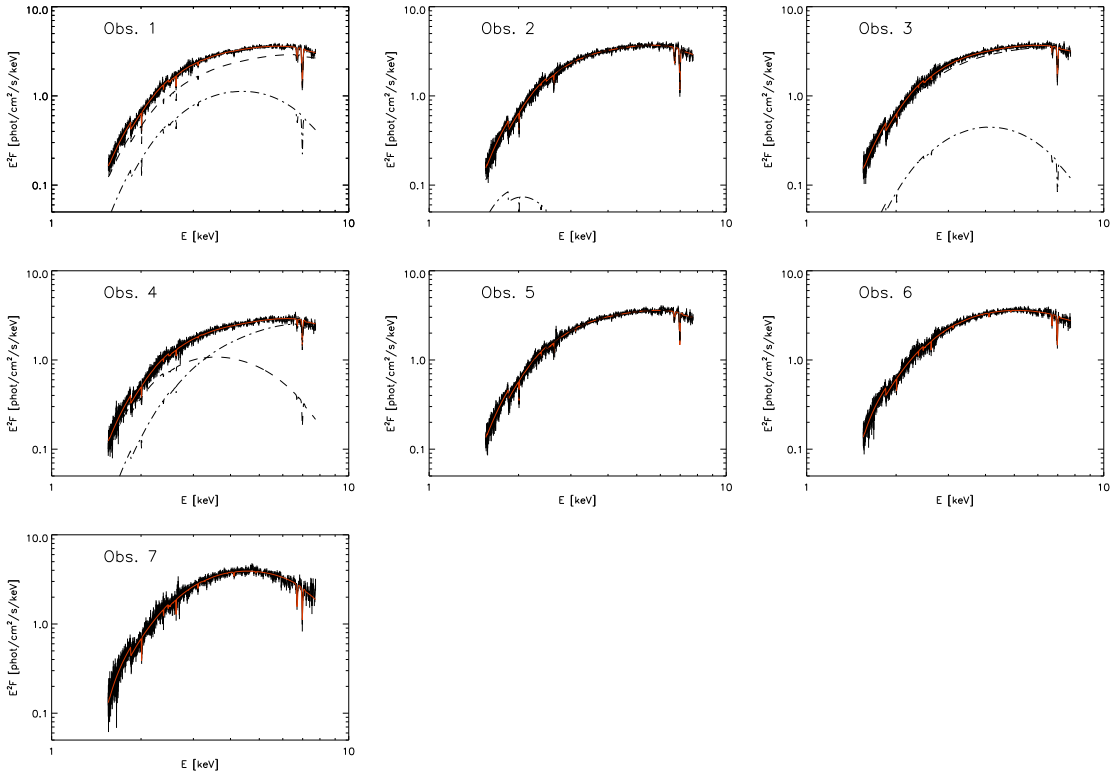}
 \caption{Unfolded HEG spectra of GX 13+1 with the model overplotted (red curve). The dashed line represents the disc black body contribution, the dashed-dotted line represents a black body contribution. The former is difficult to separate from the data in obs. 2 and obs. 5 where the contribution of the black body is negligible. In obs. 6 and 7 a {\sc spline} function is used in order to describe the continuum.} 
\vspace{-3mm}
\end{figure*}

\section{Spectral results}
\subsection{Continuum spectrum of GX 13+1}
The continuum spectrum is well described by a fit function combining an absorbed disc black body and black body model. However, we find that a different combination of continuum models such as a power-law plus a black body provides an equally acceptable fit based on the $\chi^2$ statistics. Because of the small energy coverage of an HEG spectrum and the high interstellar medium (ISM) extinction it is difficult to constrain a more physical model of the continuum. Since the goal is to characterize narrow absorption lines in the spectra the knowledge of the physical properties of the continuum are not essential. Additionally, observations 6 and 7 were performed in CC mode which has larger calibration uncertainties than the TE mode used in the rest of the observations. We find that the continuum spectra taken in the CC mode are better described using a polynomial model of 4th order, which we therefore adopt in the final fit. \\
In order to model the ISM neutral absorption we use the {\sc tbnew} model available in {\sc isis} with the \citet{Lodders2003} set of abundances. We note the presence of an absorption features around 1.8 keV (6.8 \AA). This feature could result from calibration uncertainties in the HEG instrument near the Si K-edge. \\
The spectra taken using the TE mode are significantly piled-up. The pile-up in grating spectra causes a pure reduction of the count rate. The detection of more than one photon during a readout time, which cannot be separated, is either rejected by grade selection or migrates to a higher-order spectrum \citep{Nowak2008, Hanke2009}. The pile-up effect can therefore significantly change the shape of the continuum spectrum and affect the flux measurement. We use the {\sc simple$\_$gpile3} model available in {\sc isis} in order to determine the fraction of the piled-up photons in each spectrum and correct the spectral model. In this model the pile-up fraction $\beta$ is divided by a constant representing the estimated pile-up fraction $\beta_{0}=3\times \lambda_{\rm res}\times T_{{\rm frame}}$, where $\lambda_{\rm res}$ corresponds to the absolute wavelength accuracy of the HEG spectrometer\footnotemark[1] ($\lambda_{\rm res}^{{\rm HEG}}=0.0055\ \AA$) and $T_{\rm frame}$ corresponds to the frame exposure time (1.24 s for the TE mode where one-third of the full frame is read out in the analyzed observations and 2.85 ms for the CC mode).  
The fits show that in the region between $2-4$ \AA\ the spectra taken in TE mode have the flux reduced by $\approx10-15\%$ due to the pile-up effect. The best-fit parameters are presented in the Table 2 and the best-fit model for each observation is shown in Fig. 2.\\
\citet{Diaz2012} reported the presence of an Fe K Compton broadened emission line in the {XMM-Newton} observations. However, we do not find any significant contribution from a broad emission line around  6.4 keV which could result from the fact that the effective area of the {\it Chandra} satellite drops rapidly at higher energies and is significantly smaller than that of {\sc XMM-Newton pn/MOS}. 
\subsection{Outflow properties in GX 13+1}
At first we choose the most prominent outflow lines: Fe XXVI, Fe XXV, S XVI and Si XIV. In order to determine the velocity shift and broadening of these lines we use the {\sc LineProfile} model available in {\sc isis} \citep[][Mi\v{s}kovi\v{c}ov\'{a} submitted]{Hanke2009}. In this approach we are treating each outflow line separately and do not assume they originate in the same region of the outflow.\\
The fit reveals that most of the lines are blueshifted (see Fig. 3). It seems that the velocity of the Si XIV, S XVI lines and especially Fe XXV is lower than the velocity of the Fe XXVI line. We fit a constant function to all the measured velocities of each line separately. The high value of the $\chi_{\nu}^2$ shows that the velocity of the observed absorption lines cannot be described by a constant blueshift. Additionally, we note that for the Fe lines the value of the ionic column density seems to correlate with the value of the velocity broadening in all of the observations. This correlation could indicate that the Fe lines are saturated \citep[see][]{Boirin2005}.\\
Next, we assume that all of the observed lines originate in the same regions of the outflow. We fit the absorption lines using the {\sc warmabs} model in the {\sc isis} software package which calculates the emerging spectrum after the X-ray emission has traveled through a slab of material. For comparison purposes we also use the {\sc xabs} model available in the {\sc spex} software package. In these models all ionic column densities are linked through a photoionisation model. The advantage of the {\sc warmabs}/{\sc xabs} model over the {\sc LineProfile} model is that all relevant ions are taken into account, also those which would be detected only with marginal significance using the {\sc LineProfile} model. \\
The {\sc spex} software does not provide a suitable pile-up model. Therefore, we check the influence of the pile-up on the narrow absorption lines by fitting a {\sc slab} model (which is the {\sc spex} equivalent of the {\sc LineProfile} model in {\sc isis}) to each line separately. We find that the velocity shift is consistent within the 1$\sigma$ error with the values found in the pile-up corrected spectra fitted in {\sc isis}. The ISM neutral absorption in the {\sc spex} software package is modeled using the {\sc hot} model with a very low temperature and the \citet{Lodders2009} set of abundances is used. Since in the {\sc spex} software package the spectra are not corrected for pile-up, we model the continuum spectrum using a {\sc spline} function of 4th order. \\
\begin{figure*}  
\vspace{-3mm}
\includegraphics[height=0.25\textwidth]{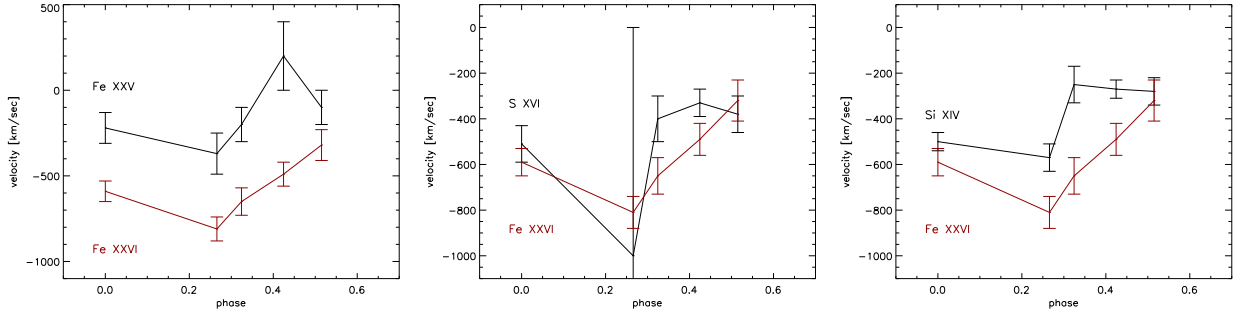}
 \caption{ The velocities of Fe XXV (left panel) S XVI (middle panel) and Si XIV (right panel) in GX 13+1 measured using the {\sc LineProfile} model in the observations $2-6$. For comparison the velocities of the Fe XXVI lines are overplotted on each panel (red color). Note that S XVI and Si XIV show on average lower velocities than Fe XXVI line. The Fe XXV line is the least blueshifted of all the analyzed lines. }
\end{figure*}

The photoionization models {\sc warmabs}/{\sc xabs} require providing an ionization balance which is calculated based on the spectral energy distribution (SED) of the source. Due to the variability in the flux and high ISM absorption towards GX 13+1 it is difficult to measure the contribution from the different energy bands and determine a proper SED. Therefore, we determine the SED based on the {\it Chandra} X-ray continuum spectrum of GX 13+1 in the energy range 0.01-12 keV. We neglect the contribution from the optical band since we do not have reliable flux measurement in this energy range. We also do not include the infra-red flux since most of these photons likely originate in the donor star and do not contribute significantly to the ionization balance of the gas in the accretion disc. Given that the absolute value of $\xi$ depends highly on the shape of the SED, only the relative changes of the $\xi$ parameter can be interpreted in this case. In order to determine the SED in each observation of GX 13+1 the continuum spectra in all observations are described by a black body and a disc black body model. The spline function which we used in order to describe the continuum in observation 6 and 7 can provide unrealistic values of the fluxes when extrapolated beyond the energy range of $1.5-8$ keV in which the spectra are fitted. We use the {\sc xstar} code when calculating the ionization balance for the {\sc warmabs} model. We provide the input density of the slab in the {\sc xstar} code of $n=10^{13}\ {\rm cm}^{-3}$, estimated for the outflow in GX 13+1 \citep{Ueda2004}, luminosity of $L=10^{38}\ {\rm erg\ s}^{-1}$ and the maximum column density of $N_{H}=10^{24}\ {\rm cm}^{-2}$. We note that close to the optically thick regime ($N_{H}\gtrapprox10^{24}\ {\rm cm}^{-2}$) the ionization balance provided by the {\sc xstar} code may not be accurate, which can affect the measurement of the column density, ionization parameter or the line width. However, this effect is negligible for the overall velocity of the outflow measured using the {\sc warmabs} model. We use the {\sc cloudy} code and the {\sc xabsinput} program when calculating the ionization balance for the {\sc xabs} model. For each $\log\xi$ between $-8.5$ and $+6.5$ with a step of $0.1$ {\sc xabsinput}\footnotemark[4]\footnotetext[4]{a detailed description of the {\sc xabsinput} program is given in the {\sc spex} manual: http://www.sron.nl/files/HEA/SPEX/manuals/manual.pdf} program calls the {\sc cloudy} code assuming a fixed density of the slab of $n=10^{8}\ {\rm cm}^{-3}$ and column density of $N_{H}=10^{16}\ {\rm cm}^{-2}$. \\
The fit to the data using the {\sc xabs} model reveals that the outflow is significantly blueshifted in all of the observations. Additionally, we indicate that the difference in the $\xi$ values provided by {\sc xabs} and {\sc warmabs} could be due to the different assumptions used when calculating the ionization balance in {\sc cloudy} and {\sc xstar} codes. We note that in the observation 1, 2 and 7 the $N_{H}$ value obtained using {\sc xabs} and {\sc warmabs} are inconsistent. The differences in the $N_{H}$ values provided by {\sc xabs} and {\sc warmabs} models could be caused by the the differences in the continuum models used in both cases and the treatment of the pile-up effect. We refit the spectra 1, 2 and 7 with the $N_{H}$ parameter fixed to the value measured using the {\sc warmabs} model and we find that the two best-fit {\sc xabs} models are statistically equivalent (99.99\% confidence level, see Table 2). Additionally, based on the 99\% joint confidence level for the column density and the ionization parameter we note that these two parameters appear correlated. This correlation is present in all of the analysed observations of GX 13+1 and regardless of the photoionization model which is used for fitting the spectra. The fact that the ionization parameter and the column density are correlated has, however, a negligible effect on the measured velocity of the outflow (see Table 2).

\subsection{Are the velocity variations of the lines caused by the orbital motion of the primary ?}
\subsubsection{GX 13+1}
Looking at the velocities measured using the {\sc warmabs}/{\sc xabs} model, it seems that the variability could in fact have an orbital nature (see Fig. 3). Hence, in this section we assume that the variations in the velocities we measure using the observations performed in 2010 are caused by the motion of the disc around the center of mass of the binary. We phase all of the observations performed in 2010 with the orbital period of 24 d \citep{corbet2010} and fit a sine function $\upsilon(\phi)=\upsilon_{0}+K\times\sin[2\pi(\phi-\phi_{0})]$ to the velocities measured using the {\sc warmabs} and {\sc xabs} models. We note after \citet{corbet2010} that the detected periodicity around $\approx24$ d appears incoherent and it could be caused by a structure that is not completely phase-locked in the binary system. Since the observations taken in 2010 show no dramatic changes in flux during the exposure (except for one dip in obs. 4 which has been excluded from the spectra), for the fit using the sinusoid function we assume the phases corresponding to the middle of each observation. Phase zero has been chosen arbitrarily and corresponds to the time when the first observation in 2010 was obtained. The fit using a sinusoid gives $\chi^2/\nu=9/2$ and $\chi^2/\nu=13/2$ for velocity measurements obtained using {\sc warmabs} and {\sc xabs} model, respectively. The large value of $\chi_{\nu}^2$ indicates that there could be systematic effects (e.g. deviations from the symmetric structure of the wind) that could influence the velocity measurement or an outlying measurement. Fig. 4 (panels a1 and b1) shows the fit and the best-fit parameters are:
$\upsilon_{0}=-430\pm30$ ${\rm km\ s}^{-1}$, $K=280\pm60$ ${\rm km\ s}^{-1}$ and $\phi_{0}=0.41\pm0.02$ for the velocity measurements obtained using the {\sc warmabs} model and $\upsilon_{0}=-390\pm20$ ${\rm km\ s}^{-1}$, $K=240\pm40$ ${\rm km\ s}^{-1}$ and $\phi_{0}=0.40\pm0.02$ for the velocity measurements obtained using the {\sc xabs} model. 
\begin{figure*} 
\vspace{-3mm}
\includegraphics[width=0.4\textwidth]{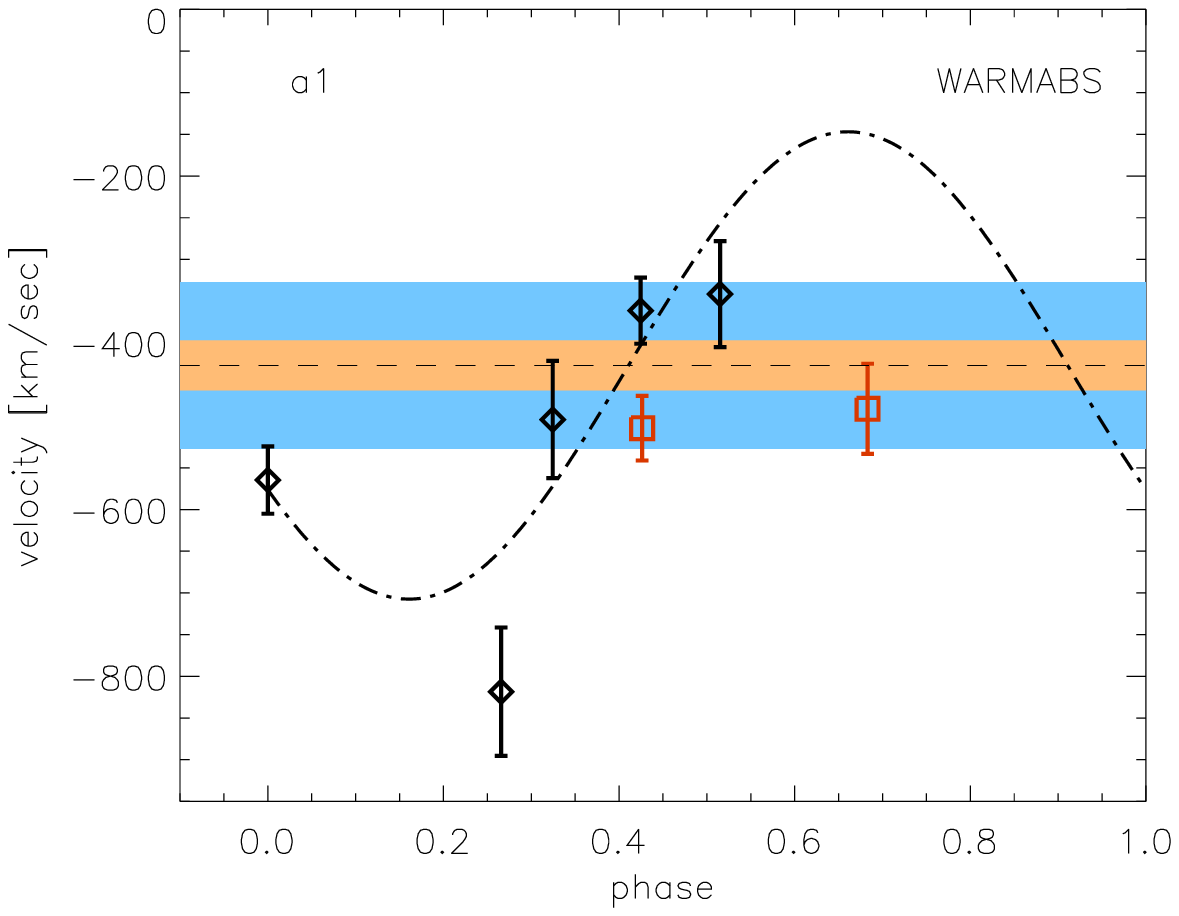}
\includegraphics[width=0.4\textwidth]{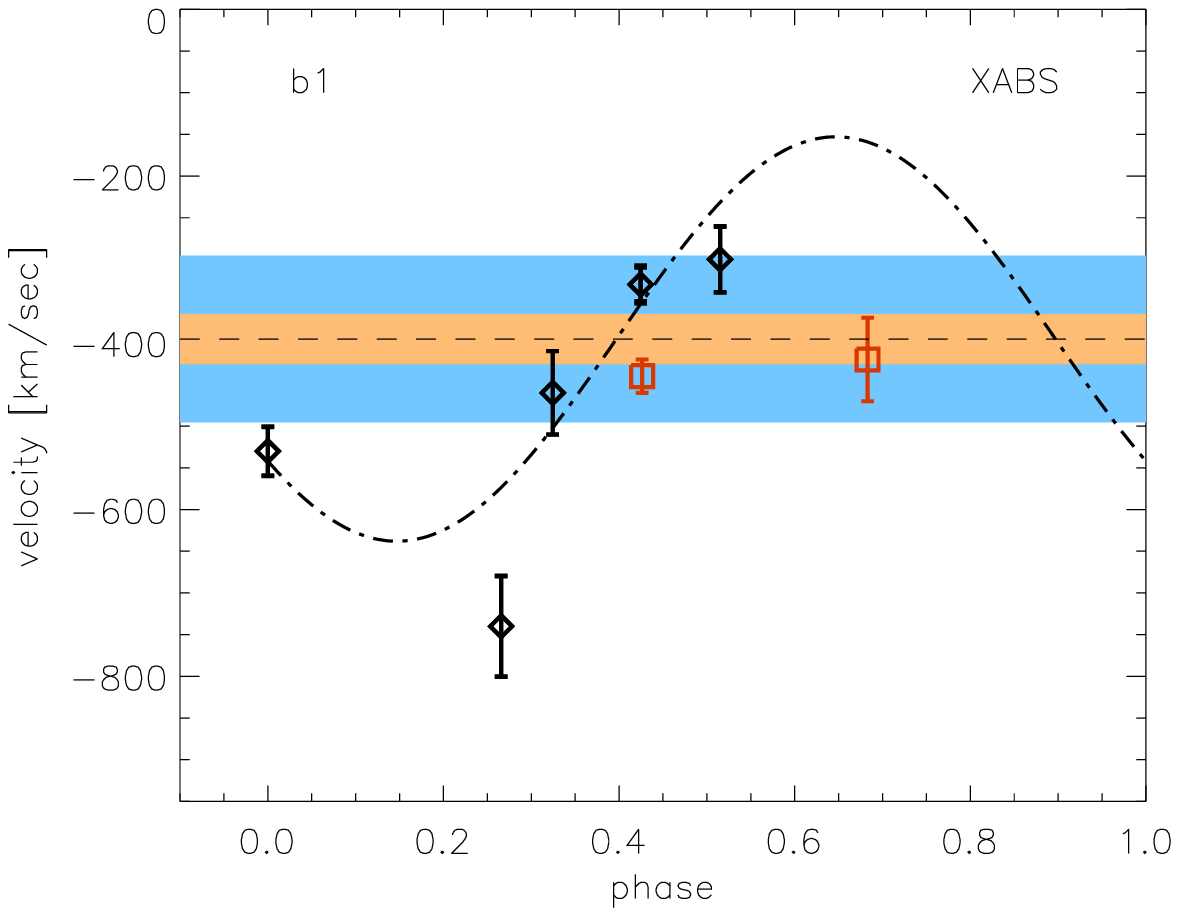}\\
\includegraphics[width=0.4\textwidth]{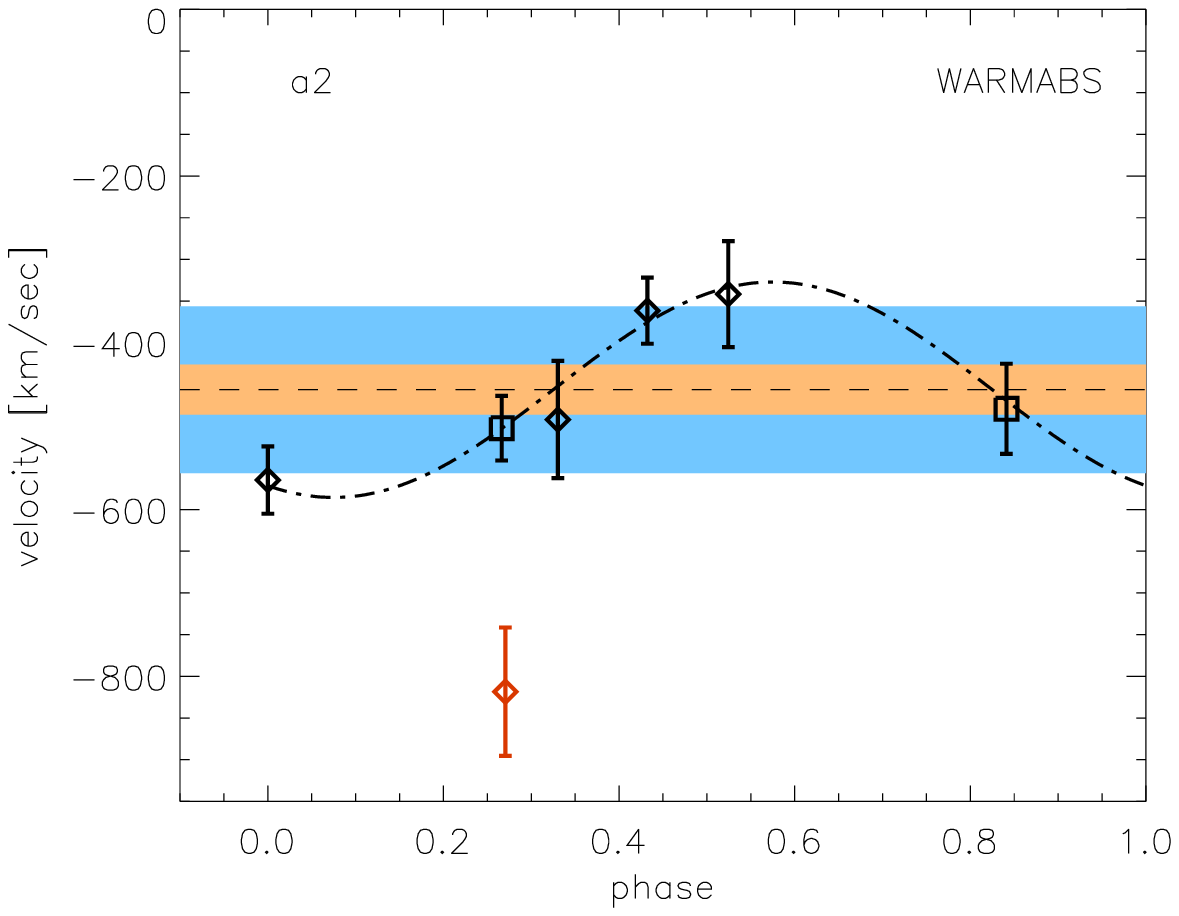}
\includegraphics[width=0.4\textwidth]{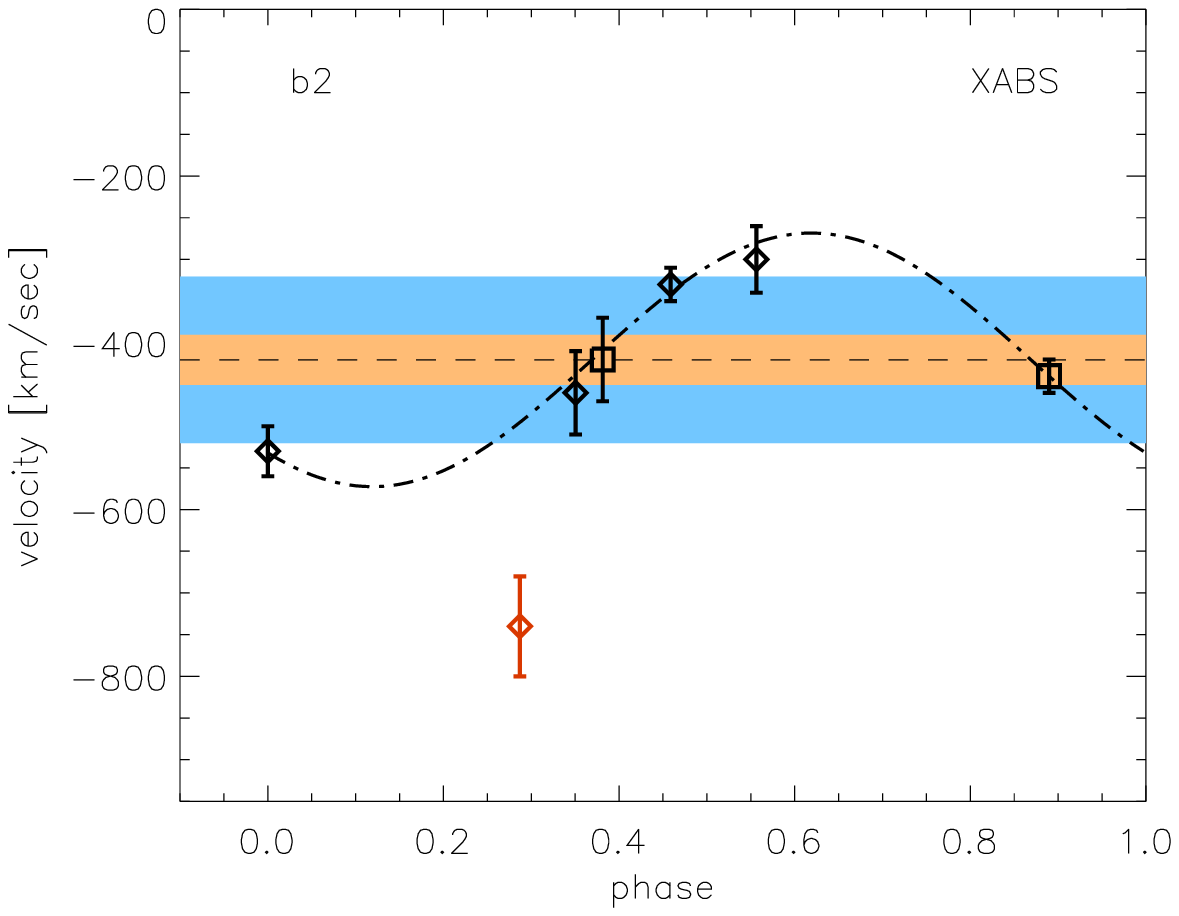}

 \caption{Velocities of all of the outflow lines measured in GX 13+1 using the {\sc warmabs} model in {\sc isis} (panels a1 and a2) and {\sc xabs} model in {\sc spex} (panels b1 and b2). The best-fit sinusoid is overplotted together with the offset $\upsilon_{0}$ determined from the fit (dashed curve). In panels a1 and b1 the orbital the phase is calculated assuming the orbital period of 24 d, whereas in panels a2 and b2 the phase is calculated assuming the orbital period of $\approx23.6$ d and $\approx22.2$ d, respectively. Additionally the observation 1 and 7 are overplotted with a square symbol and the observations excluded from the fit are marked in red. The orange and blue strip represents a predicted semi-amplitude of the radial velocity of the primary estimated assuming the donor mass of 0.8 $M_{\odot}$ and 5 $M_{\odot}$, respectively.} 
\vspace{-3mm}
\end{figure*}

\begin{table*}
\centering
\caption{Best-fit parameters obtained by fitting the model {\sc simple$\_$gpile3}$\times$[tbnew*{\sc LineProfile}*({\sc diskbb+bb})] and {\sc simple$\_$gpile3}$\times$[tbnew*{\sc warmabs}*({\sc diskbb+bb})] in the {\sc isis} package and the {\sc hot*xabs*spline} in the {\sc spex} software package to all the data of GX 13+1. The absorbed flux of the source has been corrected for pile-up only in the {\sc isis} fits. }
\begin{tabular}{l@{\,}l@{\,}c@{\,}c@{\,}c@{\,}c@{\,}c@{\,}c@{\,}c@{\,}}
\hline
\hline
Component & Parameter & 1&2&3&4&5&6&7\\
\hline
Orbital Phase &  & $-$ & 0.0  & 0.27 & 0.32 & 0.43& 0.51& $-$\\
\hline
{\it Continuum model} & & & & & & & & \\
\hline
{\sc tbnew} &  & & & & & & & \\
&$N_{\rm H}$ [$10^{22}\ {\rm cm}^{-2}$] & $4.4\pm0.1$ & $5.8\pm0.2$ & $4.75^{+0.1}_{-0.05}$ & $5.0\pm0.1$ & $5.5\pm0.2$ & $4.76\pm0.01$ &  $5.04\pm0.02$   \\
{\sc\bf diskbb} &  & & & &  & & & \\
&$T_{\rm diskbb}$ [keV]& $2.2\pm0.1$ & $1.75\pm0.02$ & $2.08\pm0.08$ & $1.04^{+0.14}_{-0.09}$ &  $1.89\pm0.02$ &$-$ & $-$\\
{\sc\bf bbody} & & & & & & & & \\
& $kT_{\rm bbody}$ [keV] & $0.98\pm0.05$ & $0.24\pm0.01$ &  $0.9\pm0.1$ & $1.54\pm0.08$ &  $0.24\pm0.03$ & $-$& $-$ \\
{\sc\bf simple\_gpile3} &  & & & & & & & \\
& $\beta_{+1}$& $1.43\pm0.02$ & $1.54\pm0.09$ & $1.3\pm0.1$ & $1.9\pm0.2$ & $1.4\pm0.1$&$-$ &$-$ \\
& $\beta_{-1}$&  $1.23\pm0.02$& $1.52\pm0.09$ & $1.3\pm0.1$ &  $1.2\pm0.2$ & $1.4\pm0.1$&$-$ &$-$ \\
\hline
{\it Outflow}          & & & & & & & & \\ 
\hline
{\sc\bf line series}  & & & & & & & & \\
 {\bf Fe XXVI} & & & & & & & & \\
& $\upsilon$  [km s$^{-1}$] & $-490\pm60$ & $-590\pm60$&  $-810\pm70$& $-650\pm80$& $-490\pm70$& $-320\pm90$& $-700\pm120$\\
& ${\upsilon^{*}_{\rm therm}}$  [km s$^{-1}$] &  $900\pm100$ & $800\pm100$&  $900\pm100$&$800\pm200$ & $120^{+520}_{-50}$& $<570$& $<140$ \\
& $ N_{\rm ion}$  [$10^{18}$ cm$^{-2}$] & $1.9\pm0.1$ & $1.8\pm0.1$ & $1.39\pm0.08$& $1.05\pm0.09$&$<50$ & $47^{+7}_{-40}$ & $50^{+10}_{-30}$\\
 {\bf Fe XXV} & & & & & & & & \\
& $\upsilon$ &  $-130\pm100$ & $-220\pm90$& $-370^{+140}_{-90}$ & $-200\pm100$&  $200\pm200$& $-100\pm100$&   $-100\pm100$\\
& ${\upsilon_{\rm therm}}$ & $1100\pm$200& $1000\pm100$ & $200\pm100$ & $< 400$ & $1000\pm300$& $<600$& $1100\pm300$\\
& $ N_{\rm ion}$  [$10^{17}$ cm$^{-2}$]          & $2.8\pm0.2$& $3.4\pm0.2$ &$ 2^{+13}_{-1}$& $4^{+18}_{-2}$& $1.8\pm0.2$& $2.4^{+33.7}_{-0.8}$&  $4.8\pm0.6$\\
 {\bf S XVI} & & & & & & & & \\
& $\upsilon$ & $-450\pm60$ & $-510\pm80$& $-1000\pm1000$& $-400\pm100$& $-330\pm60$& $-380\pm80$&  $-300\pm60$\\
& ${\upsilon_{\rm therm}}$ & $400\pm100$ & $400\pm200$&  $<13$& $<400$& $90^{+100}_{-50}$& $90^{+140}_{-60}$&  $100^{+200}_{-90}$\\
& $ N_{\rm ion}$  [$10^{16}$ cm$^{-2}$]   & $6.6\pm0.8$ &  $6\pm1$ & $7^{+18}_{-5}$ & $3.0\pm0.9$& $5^{+2}_{-1}$& $4^{+2}_{-1}$ & $8^{+27}_{-2}$  \\
 {\bf Si XIV} & & & & & & & & \\
& $\upsilon$ & $-440\pm40$&  $-500\pm40$ &  $-570\pm60$ & $-250\pm80$ & $-270\pm40$& $-280\pm60$& $-500\pm60$ \\
& ${\upsilon_{\rm therm}}$ & $370\pm70$ & $340\pm80$& $<400$&  $600\pm100$ &  $200\pm90$ & $200\pm100$&  $450\pm90$\\
& $ N_{\rm ion}$  [$10^{16}$ cm$^{-2}$]   & $7.0\pm0.6$ & $6.1\pm0.6$& $3.0\pm0.6$& $5.2\pm0.8$& $6.3\pm0.7$& $3.9\pm0.7$& $9\pm1$  \\
\hline
& $\chi^2/\nu$   & 2989/2546 & 2933/2546& 2782/2546  & 2742/2546 & 2754/2546 & 2736/2546&  2856/2546\\

\hline
\hline
{\sc\bf warmabs}  & & & & & & & & \\
& $N^{\rm warmabs}_{\rm H}$  [$10^{22}\ {\rm cm}^{-2}$] & $320\pm20$ & $170\pm10$ & $30^{+3}_{-9}$ & $70\pm30$ & $250^{+120}_{-20}$ & $130\pm50$ & $180^{+520}_{-80}$\\
& $\log \xi$ [erg cm s$^{-1}$]& $4.66^{+0.07}_{-0.01}$ & $4.43^{+0.16}_{-0.01}$ & $4.25^{+0.01}_{-0.15}$ & $4.4^{+0.1}_{-0.2}$ & $4.61^{+0.23}_{-0.02}$ & $4.6^{+0.1}_{-0.2}$ & $4.5^{+0.4}_{-0.2}$\\
& $\upsilon$ [km s$^{-1}$]& $-500\pm40$& $-560\pm40$& $-820\pm80$& $-490\pm70$& $-360\pm40$& $-340\pm60$& $-480\pm50$\\
& ${\upsilon_{\rm turb}}$ [km s$^{-1}$]& $440\pm30$ & $460\pm30$ & $780^{+100}_{-50}$ & $370\pm50$ & $320\pm30$ & $340\pm40$ & $270\pm40$\\
\hline
& $Flux_{2-10 keV}$ [$10^{-9}$ erg cm$^{-2}$ s$^{-1}$]& 6.3& 6.3& 6.4& 5.1& 6.0& 5.8& 7.2\\
\hline
& $\chi^2/\nu$ & 2919/2505 & 2880/2505 & 2757/2505  & 2727/2505 & 2713/2505 & 2700/2504&  2807/2504\\
\hline
\hline
{\sc\bf xabs}  & & & & & & & & \\
& $N^{\rm xabs}_{\rm H}$  [$10^{22}\ {\rm cm}^{-2}$] & $160\pm50$ & $50\pm20$ & $20\pm3$ & $60\pm30$ & $200\pm80$ & $80^{+60}_{-30}$ & $30\pm20$\\
& $\log \xi$ [erg cm s$^{-1}$]& $4.41^{+0.08}_{-0.04}$ & $4.2\pm0.1$ & $4.15\pm0.04$ & $4.51\pm0.02$ & $4.44\pm0.03$ & $4.35^{+0.07}_{-0.1}$ & $4.01^{+0.09}_{-0.2}$\\
& $\upsilon$ [km s$^{-1}$]& $-440\pm20$& $-530\pm30$& $-740\pm60$& $-460\pm50$& $-330\pm20$& $-300\pm40$& $-420\pm50$\\
& $\sigma_{\upsilon}$ [km s$^{-1}$]& $190\pm10$ & $200\pm10$ & $600\pm100$ & $130\pm10$ & $130\pm10$ & $140\pm10$ & $150\pm10$\\
\hline
& $\chi^2/\nu$   & 2402/2505 & 2421/2501& 2277/2500  & 2921/2500& 2187/2500 & 2135/2488&  1995/2488\\
\hline
\hline
{\sc\bf xabs$^{**}$}  & & & & & & & & \\
& $\log \xi$ [erg cm s$^{-1}$]& $4.61\pm0.02$ & $4.45\pm0.02$ & $-$ & $-$ & $-$ & $-$ & $4.45\pm0.03$\\
& $\upsilon$ [km s$^{-1}$]& $-450\pm20$& $-530\pm30$& $-$& $-$& $-$&$-$& $-410\pm40$\\
& $\sigma_{\upsilon}$ [km s$^{-1}$]& $190\pm10$ & $190\pm10$ & $-$ & $-$ & $-$ & $-$ & $120\pm20$\\
\hline
& $\chi^2/\nu$   & 2409/2506 & 2431/2502& $-$ & $-$ & $-$ & $-$ &  1997/2489\\
\hline
\hline
\end{tabular}
\\
{\footnotesize $^{*}$ Here $\upsilon_{\rm therm}^2=2kT/m_{\rm ion}+\upsilon^2_{\rm turb}$, \citep[][Mi\v{s}kovi\v{c}ov\'{a} submitted]{Hanke2009}.\\
\footnotesize $^{**}$ The $N_{H}$ value in the {\sc xabs} model is fixed to the value measured using the {\sc warmabs} model. Note that the value of $\xi$ correlate with the value of $N_H$. This correlation has, however, a negligible effect on the measured velocity of the outflow.}
\end{table*}
If these velocity variations indeed have an orbital nature the velocities measured in observations 1 and 7 should match the predicted velocities based on the fit of the sinusoid to the observations $2-6$. Given that the velocity measurement in obs. 3 deviates from the sine function fitted to the data we continue our investigation under the assumption that this measurement is an outlier and exclude it from further analysis. First, we calculate a Lomb-Scargle periodogram using the velocity measurements as a function of time in order to find periodic signals present in the data. The periodogram shows maximum power at $P\approx23.6$ d and $P\approx22.3$ d for the velocities obtained by fitting {\sc warmabs} and {\sc xabs} models, respectively. Then, we fit a sine function to all the velocity measurements of GX 13+1 obtained using the {\sc xabs} model (except obs. 3) and we let the period free during the fit. We use the period found in the Lomb-Scargle periodogram as a starting value of a period when fitting a sine function. The sinusoidal fit to the velocities obtained by {\sc warmabs} model as a function of time gives  $\chi^2/\nu=0.5/2$ and the parameters of the fit are: $\upsilon_{0}=-460\pm20$ ${\rm km\ s}^{-1}$, $K=130\pm40$ ${\rm km\ s}^{-1}$, $\phi_{0}=0.32\pm0.05$ and $P_{\rm orb}=23.57\pm0.02$ d (see Fig. 4, panel a2). 
The sinusoidal fit to the velocities obtained by {\sc xabs} model as a function of time gives  $\chi^2/\nu=0.7/2$ and the parameters of the fit are: $\upsilon_{0}=-420\pm20$ ${\rm km\ s}^{-1}$, $K=150\pm40$ ${\rm km\ s}^{-1}$, $\phi_{0}=0.37\pm0.03$ and $P_{\rm orb}=22.213\pm0.008$ d (see Fig. 4, panel b2). \\
During the refereeing process of this paper a work by \citet{Iaria2013} have been published. The authors suggest the orbital period of GX 13+1 of $P_{\rm orb}=24.5274(2)$ d. We phase all the observations except observation 3 with the period estimated by \citet{Iaria2013}. We find, however, that in this case the fit using a sine function to the velocity measurements as a function of the phase gives $\chi^2/\nu=6/3$ and $\chi^2/\nu=12/3$ for the velocities measured using the {\sc warmabs} and {\sc xabs} model, respectively. We have searched for a statistically acceptable fit around the value suggested by \citet{Iaria2013} and we obtain $\chi^2/\nu=0.6/2$ and $\chi^2/\nu=1.0/2$ with $P_{\rm orb}=25.03\pm0.01$ d for the velocities measured using the {\sc warmabs} and {\sc xabs} model, respectively.\\
In order to determine if the variability in the velocity of the absorption lines could be caused by the motion of the disc around the center of mass we calculate a semi-amplitude of the radial velocity of the primary in GX 13+1. The orbital parameters of this source are shown in Table 3. As for the mass of the donor star, several authors adopt a value of 5 $M_{\odot}$ \citep{Bandyopadhyay1999, Corbet2003}. However, mass transfer via Roche lobe overflow of a 5 $M_{\odot}$ giant on to a 1.4 $M_{\odot}$ neutron star would probably be unstable \citep{Tauris2006}. Hence, for the purpose of this work we consider also the donor star mass to be the lowest mass for a star that evolved off the main sequence during the lifetime of the Universe $M_{\rm donor}\approx0.8\ M_{\odot}$. The chosen mass of the donor star provides the lowest $K1$ velocity.
\subsubsection{H 1743$-$322, GRO J1655$-$40 and GRS 1915+105}
In the case of the sources H 1743$-$322, GRO J1655$-$40 and GRS 1915+105 we focus on the velocity of only the Fe XXVI line, given that it is the only spectral line visible in all of the analyzed spectra. Additionally, two observations of GRO J1655$-$40 and GRS 1915+105 (ObsID 5461 and ObsID 7485, respectively) show that the velocities of e.g. Fe XXVI and Fe XXV differ significantly \citep{Ueda2009, Miller2008}. In this way, we investigate whether the wind velocity in the region where the Fe XXVI line originates is stable between the observations. Additionally, we note that the outflow observed in GRS 1915+105 (ObsID 6579, 6580, 6581) shows variability (in the ionization state and column density) between the observations \citep{Neilsen2012a}. However, there is no statistically significant change in the velocity of the outflow between the observations \citep{Neilsen2012a}. We determined the velocities of the Fe XXVI line by fitting locally the {\sc slab} model. The observations ObsID 6579 and 6580 are fitted simultaneously. The continuum is described by the powerlaw and black body model absorbed by a slab of material. We fit the continuum in the positive and negative order spectra separately and couple the parameters of the $N_{H}$ and {\sc slab} absorption in the fit. The measurements agree with those published before by \citet{Miller2006a}, \citet{Miller2008}, \citet{Neilsen2012a}, \citet{Neilsen2012b} and \citet{Ueda2009}.
\begin{figure*}
\vspace{-3mm}

\includegraphics[width=0.33\textwidth]{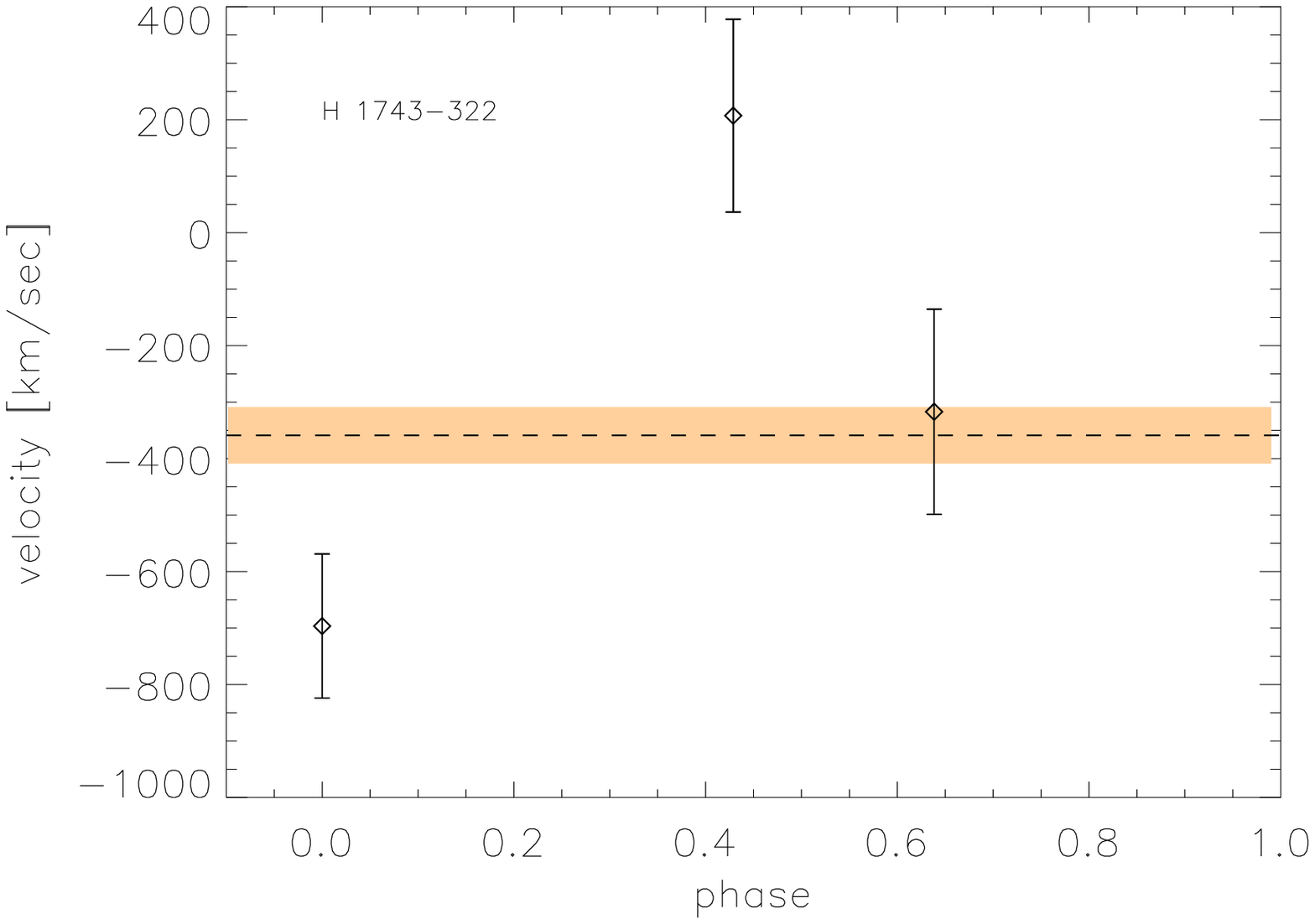}\hspace{0mm}\includegraphics[width=0.33\textwidth]{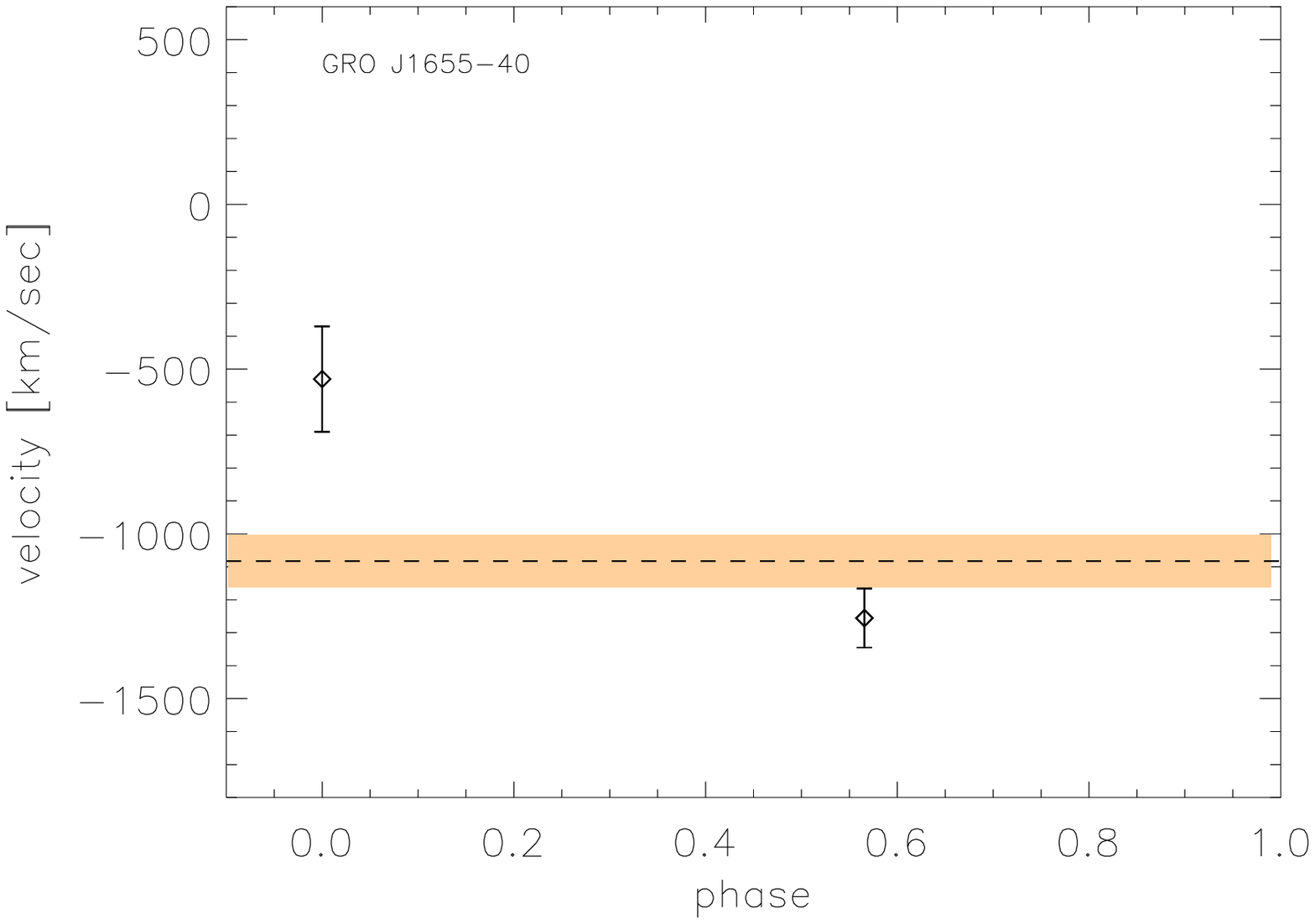}\hspace{0mm}\includegraphics[width=0.33\textwidth]{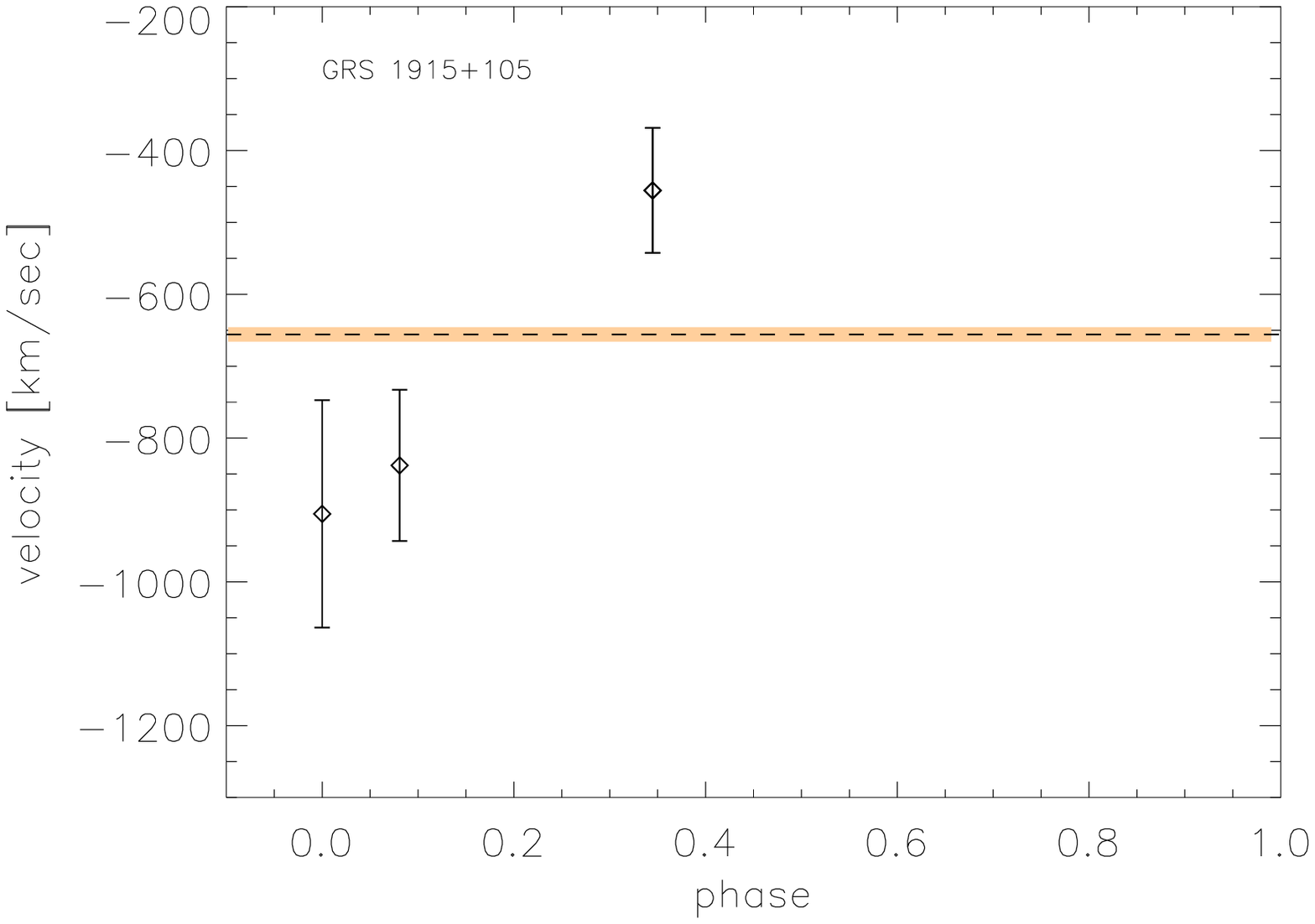}\hspace{0mm}\\

 \caption{The velocity of the Fe XXVI measured using the {\sc slab} model for H 1743$-$322 (left panel), GRO J1655$-$40 (middle panel) and GRS 1915+105 (right panel) as a function of the orbital phase. The constant offset fitted to the data is overplotted (dashed line) together with the predicted semi-amplitude of the radial velocity of the primary star (orange strip). } 
\vspace{-3mm}
\end{figure*}
\\\\
\begin{table*}
\begin{center}
\caption{Orbital parameters of LMXBs showing outflows and bound atmospheres together with the systemic velocity $\gamma$ of each binary. The approximate semi-amplitude of the radial velocity of the primary $K1$ calculated using the parameters in this table is also indicated.}
\begin{tabular}{l c c c c c c c}
\hline
\hline
 Source & $M_{BH,NS}$ [$M_{\odot}$] & $M_{\rm donor}$ [$M_{\odot}$] & $i$ [deg] & $P_{orb} $ [day] & $\gamma$ [km s$^{-1}$]  & Ref.  & $K1$ [km s$^{-1}$]\\
\hline
{\bf Outflow} & & & & & & & \\
\hline
GX 13+1 & $1.4$ & $0.8^{*}$, $5$& $60-80$ & 24 & $-70\pm 30$ & 1, 2, 3  & 30, 100\\
H 1743$-$322 & $10$ & $1$ & $60-80$ &$0.42$ &  ? & 4  & 50\\ 
GRO J1655$-$40 &  $6.29-7.60$ & $1.6-3.1$ & $63.7-70.7$ & 2.62168(14) & $-142\pm2$ & 5, 6 & 80\\
GRS 1915+105 & $10.1\pm0.6$ & $0.47\pm0.27$ &   $66\pm2$  & $33.85\pm0.16$ & $-3\pm10$ & 7, 8, 9 & 10\\
\hline 
{\bf Bound atmosphere} & & & & & &  & \\
\hline
4U 1254$-$69 & $1.20-2.64$ & $0.45-0.85$ & $65-73$ & 0.163890(9)  & $185\pm2$ & 10, 11 & 130 \\ 
MXB 1659$-$29 & $1.4$ & $0.25-0.9$ & $60-90$ & 0.296421(7) & ? & 12 & 110 \\
4U 1624$-$49 & $1.4$ & 0.8$^{*}$  & $60-90$ & 0.869907(12) & ? & 13 & 100 \\
\hline
\end{tabular} 
\end{center}
{\footnotesize 1 \citet{corbet2010}; 2 \citet{Bandyopadhyay1999}; 3 \citet{Diaz2012}; 4 \citet{Jonker2010}; 5 \citet{vanderHooft1998}; 6 \citet{Orosz1997}; 7 \citet{Steeghs2013}; 8 \citet{Fender1999}; 9 \citet{Greiner2001}; 10 \citet{Barnes2007}; 11 \citet{Motch1987}; 12 \citet{Cominsky1984}; 13 \citet{Smale2001} \\
 $^{*}$ The lowest mass for a star that evolved off the main sequence during the lifetime of the Universe.}
\end{table*}
We fit a constant to the velocities measured for each source. The fits give $\chi^2/\nu$=18/2 for H 1743$-$322, $\chi^2/\nu$=16/1 for GRO 1655$-$40 and $\chi^2/\nu$=11/2 for GRS 1915+105, respectively. Based on the high $\chi_{\nu}^2$ value for all three sources we conclude that the line velocity changes significantly between the observations. In order to determine if this variability could be caused by the motion of the disc around the binary center of mass we calculate the semi-amplitude of the radial velocity of the primary for each source. We provide the orbital parameters for all three BH LMXBs in Table 3. We note that in the case of H 1743$-$322 the orbital parameters are not well constrained. The quiescent X-ray luminosity indicates the orbital period $P>10$ h \citep{Jonker2010}. However, the infrared luminosity of this source suggests an orbital period less than the bifurcation period $P\lessapprox10-14$ h as the infrared counterpart in quiescence is too faint to be a giant, hence it is close to the main-sequence \citep{McClintock2009}. Fig. 5 shows the measured velocities, constant offset (systemic velocity of the binary + outflow velocity) and the predicted amplitude of the velocity variations based on the estimated orbital parameters of each system. The differences in the velocities of the observed lines are higher than the predicted velocity variations caused by the motion of the disc around the center of mass. 

\section{Discussion}
We have analyzed the data of GX 13+1, H 1743$-$322, GRO J1655$-$40, GRS 1915+105 and found a significant variability in the velocity of the outflow lines. The variability of the observed  absorption lines, however, appears to be higher than the predicted variability caused by the motion of the black hole/neutron star around the binary's center of mass. We discuss possible causes of this variability.\\
\subsection{The origin of the  variability in the outflow velocity}
\subsubsection{GX 13+1}
In the case of GX 13+1, the sequence of {\it Chandra} observations obtained in 2010 covered more than half of the orbital period. Hence, we had an opportunity to investigate the variability in the velocity of the outflow lines during an orbital period. The estimated $K1\approx 250$ km s$^{-1}$ based on the measured absorption line velocities would require a supergiant companion star with a mass of $\approx 50\ M_{\odot}$. This result is inconsistent with the spectroscopic and photometric measurements of the counterpart for this source suggesting that the companion is a late-type giant \citep{Bandyopadhyay1999}. \\
We find no evident correlation between the changes in the luminosity and the outflow velocity in GX 13+1. Hence, we consider a possibility that the source shows a stable outflow in which an asymmetry is introduced by e.g. a hot spot or a warp in the outer part of the disc. A hot spot in a persistent source would likely introduce a predictable asymmetric structure and show differences in the outflow characteristics (column density, ionization parameter, outflow velocity). However, any type of short lived disc warp may affect the wind in a random fashion and introduce asymmetries which are unpredictable. Although the value of column density and ionization parameter appear consistent in all observations of GX 13+1, we find evidence of degeneracy between these two parameters when fitting the outflow spectra using the {\sc xabs} and {\sc warmabs} models. Therefore, we cannot exclude a possibility that the ionization state of the outflow changes between the observations of GX 13+1. We stress that the variable density/ionization state of the source with geometrical structure of the wind unchanged is also a possible interpretation of the observed variability in the velocity.\\
By leaving out the observation showing the most blueshifted outflow lines (obs. 3) when fitting the velocity measurements using a sine function we obtain the velocity amplitude that is consistent with the radial velocity amplitude of the compact object assuming a donor star with $M\approx5\ M_{\odot}$ \citep{Bandyopadhyay1999}. Hence, the outflow velocity measurement in observation 3 could in this interpretation represent a contribution from an asymmetric structure (e.g. a hot spot or a warp) or a different ionization state of the entire wind. Additionally, we find that the velocity measurement in observations obtained in 2002 and 2011 are also consistent with the radial velocity variability of the compact object assuming the donor mass of $M_{\rm donor}\approx5\ M_{\odot}$ and taking into account the uncertainty in the orbital period of GX 13+1. Considering the binary evolution, a system with a mass ratio of $q=M_{\rm donor}/M_{NS}\gtrsim 1.5$ (transferring mass through a Roche lobe overflow) would probably lead to an unstable mass transfer and a common envelope phase \citep{Tauris2006}. The probability of observing the source in this short evolutionary stage is low. \\
The variability pattern in the outflow velocity observed in GX 13+1 looks like it could be orbital in nature. Without the full orbital coverage, however, we cannot say with certainty that the observed variations are indeed orbital. Additionally, \citet{Diaz2012} found that in the observations taken by {\it XMM-Newton} in a period of time which was smaller than the orbital period of this source, the outflow showed significant changes in its ionization state.  \\

\subsubsection{H 1743$-$322, GRO J1655$-$40, GRS 1915+105}
In the case of H 1743$-$322, the uncertainly on the orbital period makes it difficult to establish what fraction of the orbital period is covered by the presented observations. If the orbital period is indeed close to 10 h the velocity measured in each observation represents the average for the whole orbital period. Owing to the fact that it is difficult to obtain a sufficiently accurate velocity measurement when dividing these observations into phase bins, we cannot exclude the possibility that the outflow in this source could track the orbital motion of the primary. It seems, however, that on the time scale of months the wind in this source is variable beyond the variability predicted to be caused by the motion of the compact object. Given that there are significant changes in the flux between the analyzed observations \citep{Miller2006a} it is conceivable that the velocity of the wind has changed significantly between the observations.  \\
It has been suggested by \citet{Miller2006b} and \citet{Kallman2009} that the wind observed in GRO J1655$-$40 (ObsID 5461) has a high density and occurs closer to the compact object than a typical thermal wind. Hence, the authors suggest that this outflow must be driven by magnetic pressure. On the other hand, owing to the uncertainty in the density and the bolometric luminosity of this source \citet{DiazBoirin2012} suggest that the thermal/radiation pressure could still be considered as a driving mechanism for this wind. Regardless of the outflow's nature it is clear that the column density of this wind is significantly higher than typically observed in LMXBs. \citet{Zhang2012} have shown that the wind in GRO J1655$-$40 does track the motion of the compact object. However, we would like to stress that the characteristics of the wind observed in this source also change with the source luminosity. The {\it XMM-Newton} observation of GRO J1655$-$40 taken five days earlier than the discussed {\it Chandra} observation shows a wind with the column density which is significantly lower than observed in the {\it Chandra} observation \citep{Diaz2007}. Our results obtained analyzing the observations of GRO J1655$-$40 and GRS 1915+105 indicate that on a time scale of weeks/months the velocity of the Fe XXVI line does not match the predicted semi-amplitude of the radial velocity of the primary. Hence, it is possible that the wind stays stable and symmetric on a short time scales (a fraction of the orbital period) but it seems unlikely that it can remain stable for a longer time such as several orbital periods.\\
Although the method of tracking the motion of the compact object using the X-ray outflow lines \citep{Zhang2012} could be very useful in obtaining the masses of compact objects in X-ray binaries, it appears that it is difficult to find more evidence confirming the applicability of this method in the X-ray outflow observations obtained so far. The complication comes from the fact that the observations are relatively short and scattered over many years. Over such a long period of time the sources experience different spectral states which might influence the temperature and velocity of the observed outflow. In such a case it is very difficult to disentangle the intrinsic velocity of the wind from the possible Doppler shifts due to the motion of the disc. 
\subsection{Velocities of different ions in the analyzed disc outflows}
According to the simple one-dimensional thermal disc wind model proposed by \citet{Netzer2006} the ionization of the outflow becomes larger in the outer parts of the disc where the velocity of the outflow is also higher than in the inner region. In this interpretation the velocity of the Fe XXVI ions observed in the disc wind of GX 13+1 would indicate that they are located further from the compact object than the S XVI and Si XIV ions. The velocity measurements of the Fe XXVI, S XVI, Si XIV (and other absorption lines at $E < 6$ keV) observed in GRO J1655$-$40 and GRS 1915+105 seem to confirm this interpretation as well \citep{Ueda2009, Netzer2006, Kallman2009}. The velocity of the Fe XXV line, however, deviates in all observations of GX 13+1, GRO J1655$-$40 and GRS 1915+105 from the value which the model proposed by \citet{Netzer2006} would predict. \citet{Ueda2009} suggest that unlike a single electron system in H-like ions, there could be uncertainties in the atomic database for He-like (and more electrons) ions. Another possibility could be the influence of a broad Fe K emission line detected by \citet{Diaz2012} in the {\it XMM-Newton} observations of GX 13+1. The emission could potentially fill up part of the Fe XXV absorption line. Due to the low effective area of the HETGS instrument at $\approx 7$ keV it is difficult to study the influence of this emission feature on the observed absorption lines.  \\
In conclusion, the velocities of different ions in the observed thermal winds can differ significantly even if the wind is stable and symmetric which can complicate the measurement of the orbital motion of the primary further.\\
\begin{figure}  
\vspace{-3mm}

\includegraphics[width=0.5\textwidth]{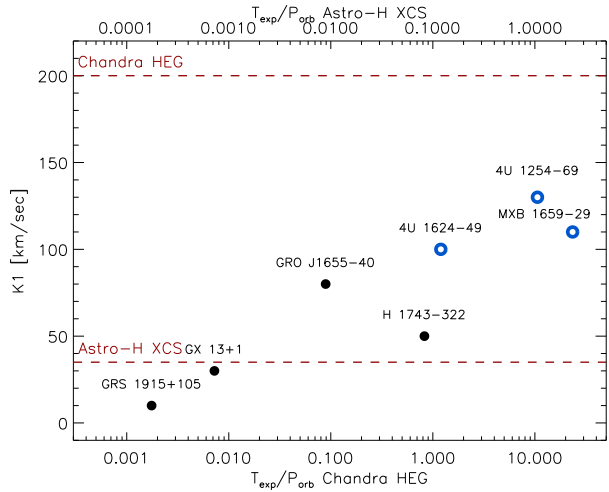}

 \caption{The x-axis indicates the estimated exposure times of the observations if taken with the {\it Chandra} HEG spectrograph as a fraction of the orbital periods of the discussed sources in order to obtain a signal-to-noise ratio of 10 per bin in the continuum. The y-axis shows the semi-amplitude of the radial velocity of the compact object estimated for each source. The spectral resolution of the {\it Chandra} HEG spectrograph and {\it Astro-H} microcalorimeter at $\approx 2$ \AA\ is indicated by the red, dashed, horizontal lines. The top axis shows the estimated exposure times as a fraction of the orbital period for the observations which could be performed by the {\it Astro-H} microcalorimeter.} 
\vspace{-3mm}
\end{figure}
\subsection{LMXBs showing bound atmospheres}
Given that the variability in the luminosity, $\xi$ and $N_{H}$ observed in thermal winds of LMXBs can affect the velocity of the outflow we propose that a better target for the method of tracking the orbital motion of the compact object would be the sources showing bound atmospheres such as 4U 1254$-$69, MXB 1659$-$29, 4U 1624$-$49. These sources have a luminosity which is too low to launch an outflow in contrast with the sample of BH/NS LMXBs analyzed in this paper. Hence, the gravitational potential in these NS LMXBs keeps the atmosphere bound to the accretion disc \citep{DiazBoirin2012}.\\
Since we have found only one good quality spectrum of these sources observed by the {\it Chandra} gratings in the archive, we briefly investigate whether it is feasible to use the absorption lines originating in the bound atmospheres in order to track the motion of the neutron stars in these systems using the current instruments. We compare the predicted $K1$ velocity and the exposure of a phase-resolved observation as a fraction of the orbital period for all sources showing outflows or bound coronae discussed in this paper. We assume a signal-to noise ratio of $\approx10$ per bin at the continuum level at $\approx7$ keV as necessary in order to obtain a precise velocity measurement for one spectral line. Given that more than one wind absorption line is usually visible in the spectra of these sources the exposure time could further be reduced by a factor of a few keeping the precision of the wind velocity measurement. We note that apart from typically observed Fe XXV and Fe XXVI absorption lines, there have been detections of the lines originating from highly ionized e.g. Ne, Mg, Si, S in the persistent emission spectra of 4U 1624$-$49 and MXB 1659$-$29 \citep{Sidoli2001,Xiang2009}.
Fig. 6 shows a comparison of the predicted $K1$ velocity based on the measured or estimated orbital parameters (see Table 3) and the exposure time of a single observation as a fraction of the orbital period. It seems that the sources showing bound atmospheres have higher $K1$ velocities than those showing outflows. However, the exposure time needed in order to obtain a good signal-to-noise ratio spectrum for the LMXBs showing bound atmospheres usually covers much more than the orbital period of the source. Hence, using current instruments it seems difficult to track the orbital motion of the compact object using the bound atmospheres present in LMXBs. Future instruments such as the microcalorimeter on {\it Astro-H} will provide a better spectral accuracy and effective area than {\it Chandra} HETGS instrument in order to track the orbital motion of the compact object in the sources showing bound atmospheres. Fig. 6 upper axis shows the predicted exposure time as a fraction of the orbital period for a observation taken by a microcalorimeter. Additionally, the approximate spectral resolution of the microcalorimeter on {\it Astro-H} is also indicated.

\section{Conclusions}

We have analyzed a sample of LMXBs showing thermal winds in order to determine whether the motion of the compact object can be tracked using the absorption lines originating in the winds driven by thermal/radiation pressure. We find that the velocity of the typical thermal wind in the analyzed sources can be variable on a time scale of months/years as well as during an orbital period. The origin of this variability could be a changing luminosity of the source that affects the wind velocity or an asymmetry in the geometry of the outflow itself. Looking at the properties of the thermal outflows we conclude that LMXBs with luminosities which are too low to launch a thermal wind but sufficient to produce a stable corona could perhaps be better targets for the method of tracking the orbital motion of the compact object. Additionally, a high column density of the observed wind/corona (like in the case of GRO J1655$-$40)  with multiple lines present in the spectrum seems to be necessary in order to measure the Doppler shifts with sufficient precision using current instruments. Alternatively, future spectrometers like the microcalorimeter on {\it Astro-H} could provide the necessary higher spectral resolution and effective area to make these measurements possible.


\begin{thebibliography}{}


\bibitem[\protect\citeauthoryear{{Bandyopadhyay}, {Shahbaz}, {Charles} \&
  {Naylor}}{{Bandyopadhyay} et~al.}{1999}]{Bandyopadhyay1999}
{Bandyopadhyay} R.~M.,  {Shahbaz} T.,  {Charles} P.~A.,    {Naylor} T.,  1999,
  \mnras, 306, 417

\bibitem[\protect\citeauthoryear{{Barnes}, {Casares}, {Cornelisse}, {Charles},
  {Steeghs}, {Hynes} \& {O'Brien}}{{Barnes} et~al.}{2007}]{Barnes2007}
{Barnes} A.~D.,  {Casares} J.,  {Cornelisse} R.,  {Charles} P.~A.,  {Steeghs}
  D.,  {Hynes} R.~I.,    {O'Brien} K.,  2007, \mnras, 380, 1182

\bibitem[\protect\citeauthoryear{{Boirin}, {M{\'e}ndez}, {D{\'{\i}}az Trigo},
  {Parmar} \& {Kaastra}}{{Boirin} et~al.}{2005}]{Boirin2005}
{Boirin} L.,  {M{\'e}ndez} M.,  {D{\'{\i}}az Trigo} M.,  {Parmar} A.~N.,
  {Kaastra} J.~S.,  2005, \aap, 436, 195

\bibitem[\protect\citeauthoryear{{Cominsky} \& {Wood}}{{Cominsky} \&
  {Wood}}{1984}]{Cominsky1984}
{Cominsky} L.~R.,  {Wood} K.~S.,  1984, \apj, 283, 765

\bibitem[\protect\citeauthoryear{{Corbet}}{{Corbet}}{2003}]{Corbet2003}
{Corbet} R.~H.~D.,  2003, \apj, 595, 1086

\bibitem[\protect\citeauthoryear{{Corbet}, {Pearlman}, {Buxton} \&
  {Levine}}{{Corbet} et~al.}{2010}]{corbet2010}
{Corbet} R.~H.~D.,  {Pearlman} A.~B.,  {Buxton} M.,    {Levine} A.~M.,  2010,
  \apj, 719, 979

\bibitem[\protect\citeauthoryear{{Diaz Trigo} \& {Boirin}}{{Diaz Trigo} \&
  {Boirin}}{2012}]{DiazBoirin2012}
{Diaz Trigo} M.,  {Boirin} L.,  2012, (arXiv:1210.0318)

\bibitem[\protect\citeauthoryear{{D{\'{\i}}az Trigo}, {Parmar}, {Miller},
  {Kuulkers} \& {Caballero-Garc{\'{\i}}a}}{{D{\'{\i}}az Trigo}
  et~al.}{2007}]{Diaz2007}
{D{\'{\i}}az Trigo} M.,  {Parmar} A.~N.,  {Miller} J.,  {Kuulkers} E.,
  {Caballero-Garc{\'{\i}}a} M.~D.,  2007, \aap, 462, 657

\bibitem[\protect\citeauthoryear{{D{\'{\i}}az Trigo}, {Sidoli}, {Boirin} \&
  {Parmar}}{{D{\'{\i}}az Trigo} et~al.}{2012}]{Diaz2012}
{D{\'{\i}}az Trigo} M.,  {Sidoli} L.,  {Boirin} L.,    {Parmar} A.~N.,  2012,
  \aap, 543, A50

\bibitem[\protect\citeauthoryear{{Fender}, {Garrington}, {McKay}, {Muxlow},
  {Pooley}, {Spencer}, {Stirling} \& {Waltman}}{{Fender}
  et~al.}{1999}]{Fender1999}
{Fender} R.~P.,  {Garrington} S.~T.,  {McKay} D.~J.,  {Muxlow} T.~W.~B.,
  {Pooley} G.~G.,  {Spencer} R.~E.,  {Stirling} A.~M.,    {Waltman} E.~B.,
  1999, \mnras, 304, 865

\bibitem[\protect\citeauthoryear{{Fleischman}}{{Fleischman}}{1985}]{Fleischman%
1985}
{Fleischman} J.~R.,  1985, \aap, 153, 106

\bibitem[\protect\citeauthoryear{{Greiner}, {Cuby} \& {McCaughrean}}{{Greiner}
  et~al.}{2001}]{Greiner2001}
{Greiner} J.,  {Cuby} J.~G.,    {McCaughrean} M.~J.,  2001, \nat, 414, 522

\bibitem[\protect\citeauthoryear{{Hanke}, {Wilms}, {Nowak}, {Pottschmidt},
  {Schulz} \& {Lee}}{{Hanke} et~al.}{2009}]{Hanke2009}
{Hanke} M.,  {Wilms} J.,  {Nowak} M.~A.,  {Pottschmidt} K.,  {Schulz} N.~S.,
  {Lee} J.~C.,  2009, \apj, 690, 330

\bibitem[\protect\citeauthoryear{{Iaria}, {Di Salvo}, {Burderi}, {Riggio},
  {D'Ai} \& {Robba}}{{Iaria} et~al.}{2013}]{Iaria2013}
{Iaria} R.,  {Di Salvo} T.,  {Burderi} L.,  {Riggio} A.,  {D'Ai} A.,    {Robba}
  N.~R.,  2013, (arXiv:1310.6628)

\bibitem[\protect\citeauthoryear{{Jonker}, {Miller-Jones}, {Homan}, {Gallo},
  {Rupen}, {Tomsick}, {Fender}, {Kaaret}, {Steeghs}, {Torres}, {Wijnands},
  {Markoff} \& {Lewin}}{{Jonker} et~al.}{2010}]{Jonker2010}
{Jonker} P.~G.,  {Miller-Jones} J.,  {Homan} J.,  {et al.},  2010, \mnras,
  401, 1255


\bibitem[\protect\citeauthoryear{{Kallman}, {Bautista}, {Goriely}, {Mendoza},
  {Miller}, {Palmeri}, {Quinet} \& {Raymond}}{{Kallman}
  et~al.}{2009}]{Kallman2009}
{Kallman} T.~R.,  {Bautista} M.~A.,  {Goriely} S.,  {Mendoza} C.,  {Miller}
  J.~M.,  {Palmeri} P.,  {Quinet} P.,    {Raymond} J.,  2009, \apj, 701, 865

\bibitem[\protect\citeauthoryear{{Lodders}}{{Lodders}}{2003}]{Lodders2003}
{Lodders} K.,  2003, \apj, 591, 1220

\bibitem[\protect\citeauthoryear{{Lodders} \& {Palme}}{{Lodders} \&
  {Palme}}{2009}]{Lodders2009}
{Lodders} K.,  {Palme} H.,  2009, Meteoritics and Planetary Science Supplement,
  72, 5154


\bibitem[\protect\citeauthoryear{{McClintock}, {Remillard}, {Rupen}, {Torres},
  {Steeghs}, {Levine} \& {Orosz}}{{McClintock} et~al.}{2009}]{McClintock2009}
{McClintock} J.~E.,  {Remillard} R.~A.,  {Rupen} M.~P.,  {Torres} M.~A.~P.,
  {Steeghs} D.,  {Levine} A.~M.,    {Orosz} J.~A.,  2009, \apj, 698, 1398

\bibitem[\protect\citeauthoryear{{Miller}, {Raymond}, {Homan}, {Fabian},
  {Steeghs}, {Wijnands}, {Rupen}, {Charles}, {van der Klis} \&
  {Lewin}}{{Miller} et~al.}{2006b}]{Miller2006a}
{Miller} J.~M.,  {Raymond} J.,  {Homan} J.,  {et al.},  2006b, \apj, 646, 394

\bibitem[\protect\citeauthoryear{{Miller}, {Raymond}, {Fabian}, {Steeghs},
  {Homan}, {Reynolds}, {van der Klis} \& {Wijnands}}{{Miller}
  et~al.}{2006a}]{Miller2006b}
{Miller} J.~M.,  {Raymond} J.,  {Fabian} A.,  {et al.} D.,  {Homan} J.,
  {Reynolds} C.,  {van der Klis} M.,    {Wijnands} R.,  2006a, \nat, 441, 953


\bibitem[\protect\citeauthoryear{{Miller}, {Raymond}, {Reynolds}, {Fabian},
  {Kallman} \& {Homan}}{{Miller} et~al.}{2008}]{Miller2008}
{Miller} J.~M.,  {Raymond} J.,  {Reynolds} C.~S.,  {Fabian} A.~C.,  {Kallman}
  T.~R.,    {Homan} J.,  2008, \apj, 680, 1359

\bibitem[\protect\citeauthoryear{{Motch}, {Pedersen}, {Courvoisier},
  {Beuermann} \& {Pakull}}{{Motch} et~al.}{1987}]{Motch1987}
{Motch} C.,  {Pedersen} H.,  {Courvoisier} T.~J.-L.,  {Beuermann} K.,
  {Pakull} M.~W.,  1987, \apj, 313, 792

\bibitem[\protect\citeauthoryear{{Neilsen} \& {Homan}}{{Neilsen} \&
  {Homan}}{2012}]{Neilsen2012b}
{Neilsen} J.,  {Homan} J.,  2012b, \apj, 750, 27

\bibitem[\protect\citeauthoryear{{Neilsen} \& {Lee}}{{Neilsen} \&
  {Lee}}{2009}]{Neilsen2009}
{Neilsen} J.,  {Lee} J.~C.,  2009, \nat, 458, 481

\bibitem[\protect\citeauthoryear{{Neilsen}, {Petschek} \& {Lee}}{{Neilsen}
  et~al.}{2012}]{Neilsen2012a}
{Neilsen} J.,  {Petschek} A.~J.,    {Lee} J.~C.,  2012, \mnras, 421, 502

\bibitem[\protect\citeauthoryear{{Netzer}}{{Netzer}}{2006}]{Netzer2006}
{Netzer} H.,  2006, \apjl, 652, L117

\bibitem[\protect\citeauthoryear{{Nowak}, {Juett}, {Homan}, {Yao}, {Wilms},
  {Schulz} \& {Canizares}}{{Nowak} et~al.}{2008}]{Nowak2008}
{Nowak} M.~A.,  {Juett} A.,  {Homan} J.,  {Yao} Y.,  {Wilms} J.,  {Schulz}
  N.~S.,    {Canizares} C.~R.,  2008, \apj, 689, 1199

\bibitem[\protect\citeauthoryear{{Orosz} \& {Bailyn}}{{Orosz} \&
  {Bailyn}}{1997}]{Orosz1997}
{Orosz} J.~A.,  {Bailyn} C.~D.,  1997, \apj, 477, 876

\bibitem[\protect\citeauthoryear{{Ponti}, {Fender}, {Begelman}, {Dunn},
  {Neilsen} \& {Coriat}}{{Ponti} et~al.}{2012}]{Ponti2012}
{Ponti} G.,  {Fender} R.~P.,  {Begelman} M.~C.,  {Dunn} R.~J.~H.,  {Neilsen}
  J.,    {Coriat} M.,  2012, \mnras, 422, L11

\bibitem[\protect\citeauthoryear{{Ratti}, {Steeghs}, {Jonker}, {Torres},
  {Bassa} \& {Verbunt}}{{Ratti} et~al.}{2012}]{Ratti2012}
{Ratti} E.~M.,  {Steeghs} D.~T.~H.,  {Jonker} P.~G.,  {Torres} M.~A.~P.,
  {Bassa} C.~G.,    {Verbunt} F.,  2012, \mnras, 420, 75

\bibitem[\protect\citeauthoryear{{Sidoli}, {Oosterbroek}, {Parmar}, {Lumb} \&
  {Erd}}{{Sidoli} et~al.}{2001}]{Sidoli2001}
{Sidoli} L.,  {Oosterbroek} T.,  {Parmar} A.~N.,  {Lumb} D.,    {Erd} C.,
  2001, \aap, 379, 540

\bibitem[\protect\citeauthoryear{{Smale}, {Church} \&
  {Ba{\l}uci{\'n}ska-Church}}{{Smale} et~al.}{2001}]{Smale2001}
{Smale} A.~P.,  {Church} M.~J.,    {Ba{\l}uci{\'n}ska-Church} M.,  2001, \apj,
  550, 962

\bibitem[\protect\citeauthoryear{{Soria}, {Wickramasinghe}, {Hunstead} \&
  {Wu}}{{Soria} et~al.}{1998}]{Soria1998}
{Soria} R.,  {Wickramasinghe} D.~T.,  {Hunstead} R.~W.,    {Wu} K.,  1998,
  \apjl, 495, L95

\bibitem[\protect\citeauthoryear{{Steeghs}, {McClintock}, {Parsons}, {Reid},
  {Littlefair} \& {Dhillon}}{{Steeghs} et~al.}{2013}]{Steeghs2013}
{Steeghs} D.,  {McClintock} J.~E.,  {Parsons} S.~G.,  {Reid} M.~J.,
  {Littlefair} S.,    {Dhillon} V.~S.,  2013, \apj, 768, 185

\bibitem[\protect\citeauthoryear{{Tauris} \& {van den Heuvel}}{{Tauris} \& {van
  den Heuvel}}{2006}]{Tauris2006}
{Tauris} T.~M.,  {van den Heuvel} E.~P.~J.,  2006, {Formation and evolution of
  compact stellar X-ray sources}.
pp 623--665

\bibitem[\protect\citeauthoryear{{Ueda}, {Murakami}, {Yamaoka}, {Dotani} \&
  {Ebisawa}}{{Ueda} et~al.}{2004}]{Ueda2004}
{Ueda} Y.,  {Murakami} H.,  {Yamaoka} K.,  {Dotani} T.,    {Ebisawa} K.,  2004,
  \apj, 609, 325

\bibitem[\protect\citeauthoryear{{Ueda}, {Yamaoka} \& {Remillard}}{{Ueda}
  et~al.}{2009}]{Ueda2009}
{Ueda} Y.,  {Yamaoka} K.,    {Remillard} R.,  2009, \apj, 695, 888

\bibitem[\protect\citeauthoryear{{van der Hooft}, {Heemskerk}, {Alberts} \&
  {van Paradijs}}{{van der Hooft} et~al.}{1998}]{vanderHooft1998}
{van der Hooft} F.,  {Heemskerk} M.~H.~M.,  {Alberts} F.,    {van Paradijs} J.,
   1998, \aap, 329, 538

\bibitem[\protect\citeauthoryear{{Xiang}, {Lee}, {Nowak}, {Wilms} \&
  {Schulz}}{{Xiang} et~al.}{2009}]{Xiang2009}
{Xiang} J.,  {Lee} J.~C.,  {Nowak} M.~A.,  {Wilms} J.,    {Schulz} N.~S.,
  2009, \apj, 701, 984

\bibitem[\protect\citeauthoryear{{Zhang}, {Liao} \& {Yao}}{{Zhang}
  et~al.}{2012}]{Zhang2012}
{Zhang} S.-N.,  {Liao} J.,    {Yao} Y.,  2012, \mnras, 421, 3550



\end{thebibliography}

\section{Acknowledgements}
OKM thanks Mafred Hanke, Thomas Dauser, Rob Detmers, Jelle Kaastra, Tullio Bagnoli and Ton Raassen for useful discussions and help with the software packages for spectral fitting. We thank the referee for useful comments that helped us improve the paper. This research has made use of a collection of {\sc isis} scripts provided by the Dr. Karl Remeis observatory, Bamberg, Germany at http://www.sternwarte.uni-erlangen.de/git.public/?p=isisscripts. IM acknowledges funding from the European Commission under grant agreement ITN 215212 `Black Hole Universe' and partial support by GACR grants 13-33324S and 3-39464J.

\end{document}